\documentclass[twocolumn]{aastex62}
\usepackage[utf8]{inputenc}
\usepackage{amsmath}
\usepackage{rotating}
\usepackage{comment}
\graphicspath{{./}{figures/}}

\shorttitle{Spitzer Phase Curves}
\shortauthors{Swain et al.}

\begin{document}
\title{Thermal Phase Curves in Hot Gas Giant Exoplanets Exhibit a Complex Dependence on Planetary Properties}


\correspondingauthor{Mark R. Swain}
\email{mark.r.swain@jpl.nasa.gov}

\author[0009-0001-4487-7299]{Mark R. Swain}
\affil{California Institute of Technology, NASA Jet Propulsion Laboratory }

\author[0000-0002-5785-9073]{Kyle A. Pearson}
\affil{California Institute of Technology, NASA Jet Propulsion Laboratory }

\author[0000-0002-9258-5311]{Thaddeus D. Komacek}
\affil{University of Maryland, Department of Astronomy}

\author[0000-0001-5966-837X]{Geoffrey Bryden}
\affil{California Institute of Technology, NASA Jet Propulsion Laboratory }

\author[0000-0002-3099-0493]{Emeline Fromont}
\affil{University of Maryland, Department of Astronomy}

\author[0000-0000-0000-0000]{Gautam Vasisht}
\affil{California Institute of Technology, NASA Jet Propulsion Laboratory }

\author[0000-0002-7402-7797]{Gael Roudier}
\affil{California Institute of Technology, NASA Jet Propulsion Laboratory }

\author[0000-0012-3245-1234]{Robert T. Zellem}
\affil{Goddard Space Flight Center }



\begin{abstract}
We present a catalog of uniformly processed 3.6-$\mu$m and 4.5-$\mu$m band exoplanet thermal phase curves based on Infrared Array Camera observations obtained from the Spitzer Heritage Archive. The catalog includes phase curve measurements for 34 planets, 16 of which contain full orbit coverage and have detectable secondary eclipses in both channels. The data are processed in the EXCALIBUR pipeline using a uniform analysis consisting of aperture photometry and modeling of instrument effects along with the exoplanet signal. Nearest-neighbors regression with a Gaussian kernel is used to correct for instrumental systematics correlated to the star's centroid position and shape in conjunction with a novel test to avoid overfitting. These methods may have utility in addressing sub-pixel gain variations present in modern infrared detectors. We analyze the 3.6-$\mu$m and 4.5-$\mu$m phase curve properties and find a strong wavelength-dependent difference in how the properties correlate with physical parameters as well as evidence that the phase curve properties are determined by multiple physical parameters. We suggest that differences between the 3.6-$\mu$m and 4.5-$\mu$m phase curve properties are due to 3.6~$\mu$m observations probing regions of the atmosphere which could include a cloud layer. Taken together, the observed phase curve behavior suggests that different physical processes are responsible for establishing the thermal phase curve at different pressures, which are probed by different wavelengths, and that further 3D GCM modeling is required to investigate the reason for this complex dependence on planetary properties.
\end{abstract}


\section{Introduction} \label{sec:intro}

The Spitzer Space Telescope was used for some of the earliest observations of infrared emission from exoplanet atmospheres (\citealt{Charbonneau2005}; \citealt{Charbonneau2008}; \citealt{Barman2008}; \citealt{Knutson2008}; \citealt{Knutson2009}; for a comprehensive review see \citealt{Deming:2020aa}). Exoplanets with short orbital periods are close enough to their host star to reach hundreds or thousands of degrees Kelvin making them bright thermal sources and excellent targets for Spitzer. During an eclipse, when a planet passes behind the parent star, the star blocks out flux coming from the planet making it possible to establish the stellar contribution and thus infer a baseline for the thermal emission from the planet emitted during the planet's orbit. By observing thermal emission during the entire exoplanet orbit, termed the phase curve, it is possible to gain insights into the properties of the dayside and nightside atmospheres of these distant worlds. Thermal emission is highly sensitive to both the temperature profile and the vertically integrated column abundances and thus it is highly desirable to have phase curve measurements at multiple wavelengths \citep{Heng:2014b,Showman:2020rev,Zhang:2020rev}. 

\cite{showman2002} predicted that close-in giant planets should be characterized by significant day-night photospheric temperature variations, with the possibility of an asymmetry in the light curve due to heat transport by zonal winds. Both predictions were qualitatively verified by observations in the infrared \citep{harrington2006,knutson2007}. For tidally-locked hot-Jupiter type planets, the observed flux difference between day side and night side is believed to be caused by a combination of temperature changes and, potentially, differences in atmospheric composition and cloud coverage that, in turn, modulate the opacity. Recent work has investigated several factors such as the dissociation of molecular hydrogen \citep{Tan:2019aa,Roth:2021un,jacobs2022}, atmospheric drag in the presence of nightside clouds \citep{Roman:2021wl,Parmentier:2021tt,tan2024}, and planetary rotation period \citep{Rauscher:2014,roth2024} that influence the dayside to nightside heat transport process.

Here we report on the analysis of a catalog of uniformly analyzed Spitzer 3.6- and 4.5-$\mu$m phase curves (see Appendix Table~\ref{tab:phasecurve_catalog_pid} program ID and related information for each target). Previous studies of the properties of Spitzer phase curves \citep{schwartz2015,schwartz2017,zhang2018,beatty2019,keating2019,may2020,may2021,may2022} have been reported including two employing uniform analysis methods \citep{bell2021,dang2024}. What makes our study unique is the combination of a uniform analysis method and the largest sample of planets to date with measured phase curve properties for both of the Spitzer 3.6- and 4.5-$\mu$m IRAC channels. Another difference between our work and previous results is how a uniform analysis is implemented. The previous uniform analyses \citep{bell2021,dang2024} implemented a uniformly applied metric to select between different approaches for modeling the instrument systematic errors. In our study, a single method for modeling the instrument systematic errors is applied to every phase curve. Our approach to the sub-pixel gain correction may be useful for exoplanet time series observations with modern HgCdTe detectors where small, sub-pixel gain variation have been measured \citep{shapiro2018}. 

Our study builds on the important body of work by substantially enlarging the sample of planets  having a uniform data reduction and analysis method applied to both the 3.6-$\mu$m and 4.5-$\mu$m Spitzer phase curves. This work includes the analysis of multiple phase curve visits when those observations exist. There are numerous lines of inquiry supported by our catalog that are beyond the science focus of this manuscript and the data products from our study are available to the community. We also introduce standardized methods to explicitly check for residual correlated noise and for instrument model over-fitting of the data. We then use the measured properties of phase curves from the catalog and we report on comparing these observational results to Global Circulation Model (GCM) predictions. We also characterize the trends in phase curve model harmonics, which may provide additional information about the phase curve properties.

\section{Observations} \label{sec:obs}

Targets are selected from the Spitzer Heritage Archive\footnote{https://sha.ipac.caltech.edu/applications/Spitzer/SHA/} because they have publicly available observations taken with the Infrared Array Camera (IRAC) over the duration of the planets' full orbit \citep{Fazio2004}. Images are processed starting with the Basic Calibrated Data (BCD) at the native resolution in both full-frame and sub-frame mode. The sub-frame mode consists of a datacube with 64 frames each containing a 32 $\times$ 32 pixel image ($39'' \times 39''$). The full frame mode is 64 times larger than the sub-frame (256 $\times$ 256) and has the same plate scale at $1.25''$/pixel. The time of each frame is assumed to occur uniformly between the start and end of the integration. The header keyword MBJD OBS (start of the first image), AINTBEEG (integration between) and ANTIMEEND (integration end) are used to compute the time of each frame. Aperture photometry is used to extract the flux from each image. The stellar centroid is estimated from a flux weighted position in a 5$\times$5 grid around the target coordinate based on the WCS header information. Multiple aperture sizes are used, ranging from 2--4 pixels at steps of 0.1 pixel. Ultimately, the aperture which minimizes the scatter in the photometric time series is used for the light curve fitting. A background subtraction is done using the median of an {optimal} annulus tested between 7 and 15 pixels from the target in steps of 1 pixel. We also found in some cases, the background annulus can be contaminated by crowded fields so we perform an additional masking for every pixel brighter than the 95\textsuperscript{th} percentile in the annulus to remove bright background sources that would skew the estimate. The noise pixel parameter is an estimate for the width of the point spread function on the detector. For more information on the noise pixel parameter, see section 2.2.2 of the IRAC instrument handbook or the Appendix A of \citet{Lewis2013}. The noise pixel often correlates to systematic changes in measured brightness and is quicker to estimate than fitting a 2D Gaussian to each frame, which is why we use it as a feature in our instrument model. In addition to the noise pixel parameter, we also use a flux weighted centroid in our instrument model described in the next section. Individual phase curves require anywhere from a day to a week of continuous data and contain typically over 10,000 images for full-frame data and 1--2 orders of magnitude more for the sub-frame mode. 

In our analysis, we opted to use the noise pixel parameter and centroid position for detrending, rather than fitting a full 2D Gaussian PSF model as employed in several previous studies (e.g., Mendonca et al. 2018, Lanotte et al. 2014, Demory et al. 2016a, 2016b). This choice was primarily driven by computational efficiency, as PSF fitting significantly increased run times by 1--2 orders of magnitude in our tests. Given the large number of frames in our phase curve observations, this additional computational cost was prohibitive for our multi-system analysis. While we acknowledge that a direct comparison between our approach and full PSF fitting would provide valuable validation, such an extensive reanalysis was beyond the scope of this work. Based on our experience with this dataset, we found that in approximately half of our targets, the noise pixel parameter was not necessary to adequately model the systematics, with most correlated noise captured by the centroid position alone. We believe our simplified model achieves a reasonable balance between computational efficiency and systematic noise mitigation, enabling analysis of a larger sample under consistent modeling assumptions. However, we recognize this as a potential limitation of our approach, and future work directly comparing these methods could help quantify any differences.

\section{Methods}

\begin{figure}[h]
\includegraphics[scale=0.34]{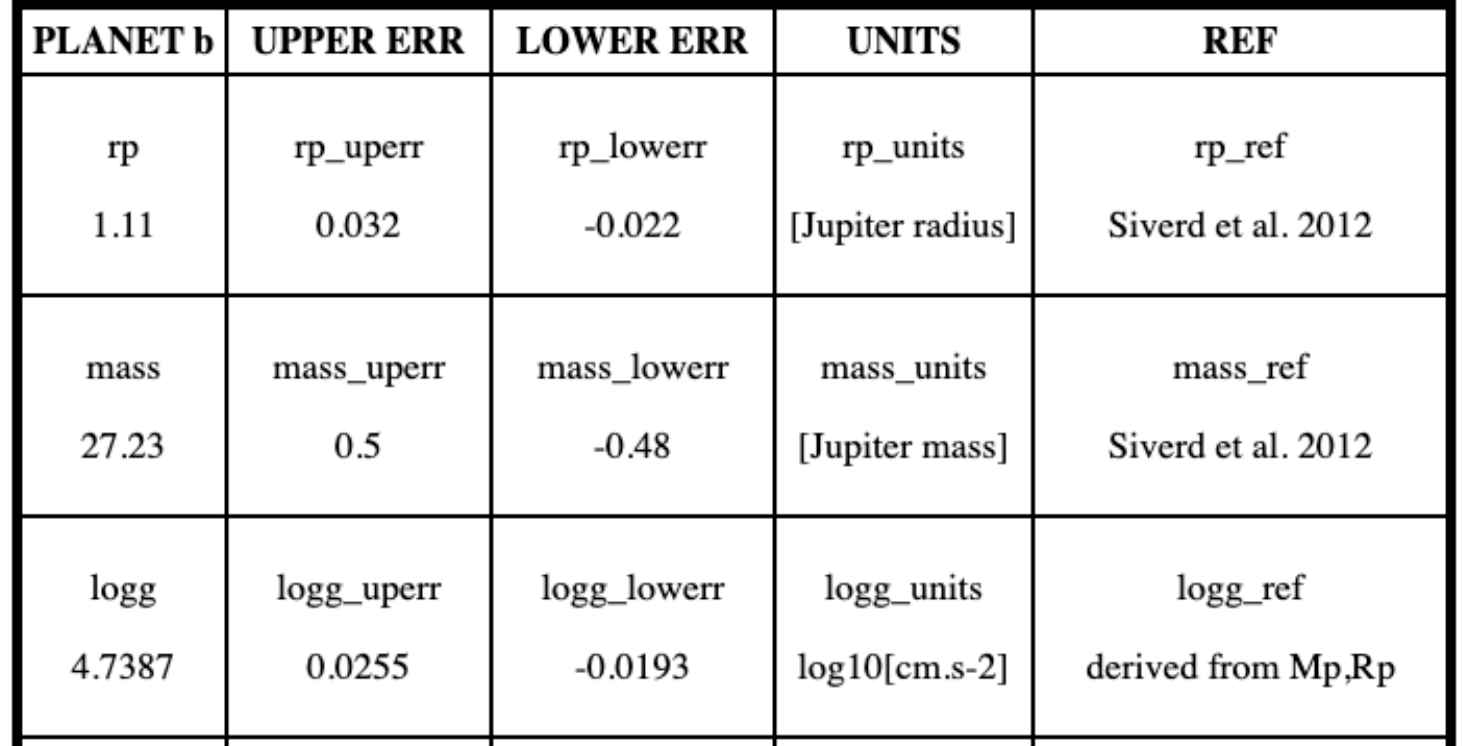}
\hspace{-1in}
\caption{Selected planet parameters for KELT-1 b used in this analysis provided an example of the kind of information available for each planet in our catalog from the IPAC EXCALIBUR portal using the PlanetName.system.finalize.parameters data product. The full planet parameters table available online includes values for radius, mass, log(g), equilibrium temperature, semi-major axis, period, time of mid-transit, inclination, eccentricity, omega, impact parameter, transit duration, and transit depth. An exoplanet host star parameter table following a similar format is also included in the data product.}
\label{planet_parameters}
\end{figure}

\begin{figure*}[t]
\hspace{-0.25in} \\
\centering
\includegraphics[scale=0.4]{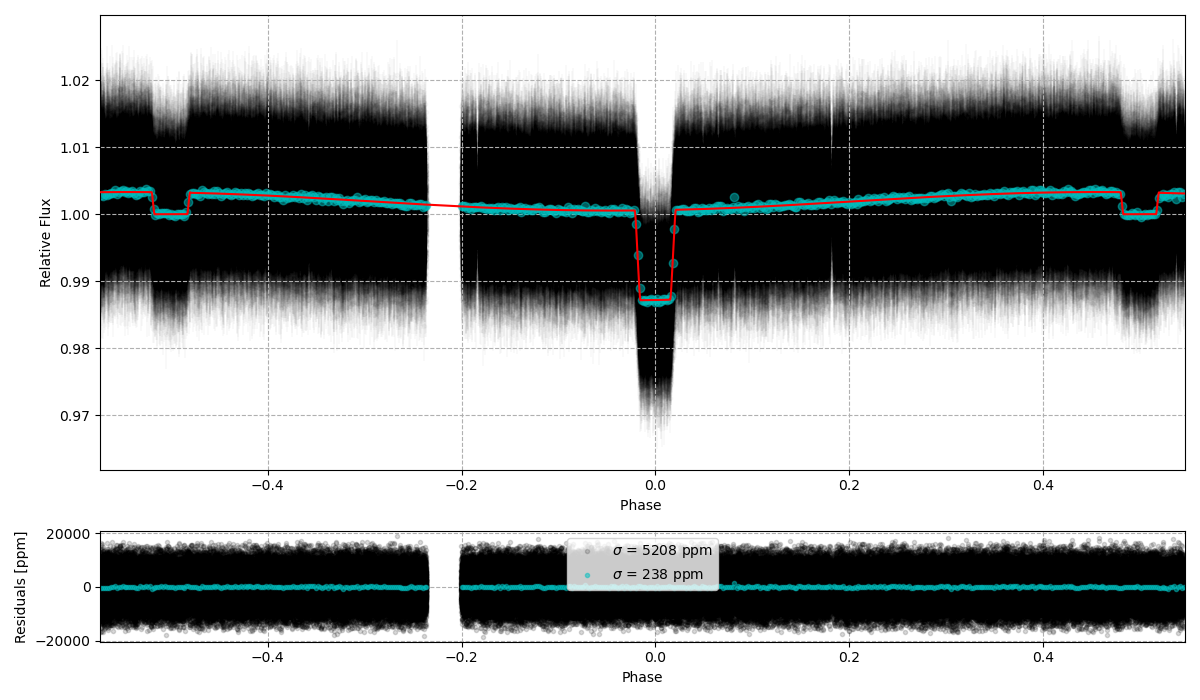}\\
\caption{The phase curve for KELT-20 b from IRAC 4.5-$\mu$m channel in units of relative flux after applying the instrument model (top) and the residuals after subtracting the astrophysical and instrument models (bottom). These plots are available as part of the associated data products for all the planets in this catalog. The black data points are the observations at the native resolution after applying the pixel map to detrend for systematics. The colored data points are binned to a resolution of 1 minute purely for visualization purposes. The red line indicates the best-fit phase curve model to the data.} 
\label{phasecurve}
\end{figure*}

This uniform analysis of Spitzer phase curve observations is implemented in the EXCALIBUR (EXoplanet CALIbration and Bayesian Unified Retrieval) science data pipeline and represents an expansion on previously reported capability \citep{swain2021,roudier2021,huber-feely2022}. EXCALIBUR is designed to enable comparative planetology through uniform processing of an input catalog. EXCALIBUR maintains the chain of inference for a specific processing instance through persistent data products (see \citet{swain2021} for further description). The final and intermediate data products associated with the Spitzer phase curve catalog reported in this manuscript are publicly available via the EXCALIBUR portal maintained at https://excalibur.ipac.caltech.edu/Excalibur/. The available data products are described in the EXCALIBUR Users Guide document also available on the EXCALIBUR portal. Throughout the manuscript, we identify the specific EXCALIBUR data products that contain information such as input parameters, intermediate results, and final results (for examples, see Figures~\ref{planet_parameters},\ref{phasecurve},\ref{spitzer_posteriors}).

\subsection{System Light Curves}

The system light curves, also termed phase curves, are fit simultaneously with an instrument model and astrophysical model. The instrument model characterizes systematics that are correlated to stellar position and the width of the star's PSF as measured with the noise pixel metric (e.g., \citealt{Ingalls2012}; \citealt{Zellem2014}; \citealt{Ingalls2016}). The astrophysical model characterizes the size of the planet, time of transit, eclipse depth, and phase curve variations. The astrophysical signal and instrument model are applied in a two-step process where the raw data is first detrended of the astrophysical signal and then variations are detrended with respect to systematic parameters using a nearest-neighbors approach and a Gaussian kernel. The instrument model uses nearest-neighbor regression with respect to centroid position, noise pixel, and a time-dependent component that correlates data subject to a ``ramp'' effect. Occasionally, there are gaps in the observing sequence spanning minutes to hours when the spacecraft turns to downlink data back to Earth. When the telescope repositions, the instrument displays a photometric effect that causes the target to change brightness over time on the order $\sim1$\%, which decays over the time span of a few hours. Modeling the ramp effect with position-dependent systematics is done simultaneously in our instrument model using nearest neighbor regression of key metrics or features. The instrument model uses a flux weighted centroid, the noise pixel parameter and an exponentially decaying feature that uses time-since-gap as an input with an e-folding time of 1 hour. We opt for a semi-time-dependent model whereby the exponential feature is set to 0 after 3 hours allowing for a more traditional instrument based on PSF shape to take over. This approach is contrary to past literature that usually fits each ramp section individually \citep{deming2006,knutson2007,Agol2010,Zellem2014}, thereby adding to the number of free parameters, which we found to ultimately hurt the analysis of sensitive phase curve signals by opening up the possibility for degeneracies while significantly increasing the model-fitting run time. Instead, we correlate the ramp section simultaneously with our PSF data which simplifies the retrieval of subtle features. The {\bf observed flux} is modeled simultaneously using,

\begin{equation}
    F_{obs} = F_{astrophysical}*F_{instrument}.
\end{equation}
\noindent
Here $F_{obs}$ is the flux recorded on the detector, $F_{astrophysical}$ is the astrophysical signal (i.e., the transit light curve or eclipse) and $F_{instrument}$ is a residual correction factor. The specifics of each model are discussed in the subsections below. Nested sampling is used to infer the final parameters and uncertainties (see Figure~\ref{spitzer_posteriors}) for the astrophysical model because it speeds up Bayesian inference.

\subsection{Instrument Model}

Motions on board the spacecraft, for example a reaction wheel being used for guiding, can translate into sub-pixel motions that result in systematic variations correlated to position and, in some cases, time (i.e., the ``ramp'' effect). For each data set, an instrument model is constructed using nearest-neighbors regression with a Gaussian kernel. This technique is also known as the pixel-map (e.g., \citealt{Ballard2010}; \citealt{Lewis2013}; \citealt{Zellem2014}) and the corrections can be decomposed back into a something analogous to a sub-pixel flat field (see lower panels in Figure~\ref{36vs45}). The instrument response is estimated after an astrophysical signal has been divided out of the time series. This prevents the planetary signal from interfering with the correction. However, a caveat for this approach is that the instrument model can absorb some error in a poor fit during the detrending step, thus leading to posteriors larger than a disjointed approach would, that detrends the data prior to fitting. A Gaussian kernel refers to a weighted average of $N$ of the nearest neighbors, where we search for neighbors in 4 of our instrument features/parameters, $p$, ramp-time, noise pixel parameter, and flux weighted centroids: x, y centroid and noise pixel. An individual correction for the $i$\textsuperscript{th} data point is formulated as follows: 

\begin{equation}
 A_{i} = \sum^{}_n \prod^{}_{p} exp\left( \frac{-w(X_{n,p}-\bar{X}_{i,p})^{2}}{2\sigma^2_{p}}\right)
\end{equation}

\begin{equation} \label{instrument}
    F_{instrument},i = \sum^{}_n \frac{F_{n}}{A_{i}} \prod^{}_{p} exp\left( \frac{-w_p(X_{n,p}-\bar{X}_{i,p})^{2}}{2\sigma^2_{p}}\right)
\end{equation}
where $A_i$ defines a normalization term such that the sum of the Gaussian weights is unity and $F_i$ is the flux from the time series after an astrophysical signal is divided out.  The standard deviation of each parameter, $\sigma_p$, is computed for each set of $N$ neighbors. The number of neighbors depends on a timescale rather than an arbitrary population size. For a given  data set, the number of neighbors equals the number of observations occurring within a 10-minute range up to a maximum value of 500 nearest neighbors. Data sets taken in full-frame mode tend to have longer cadences and will often have neighbors in the 50--100 range, whereas sub-frame data sets tend to have a few hundred nearest neighbors. We tested various time scales and found the more neighbors that are used, the weaker the instrument model behaves as a result of averaging over high frequency variations. \cite{may2020} found biases in best fit values due to degeneracies between astrophysical and systematic models when binning data in time before modeling. Therefore, we adopt a similar approach and process everything at the native resolution, with each point having its own set of weighted neighbors, enabling sensitivity to both rapid variations and longer-term trends. The chosen timescale balances the capture of local systematic effects without over-smoothing high-frequency signals. For bright targets in subframe mode with short exposures, this typically results in effective timescales of a few minutes, while full-frame mode observations generally use 50--100 neighbors. The robustness of this approach is validated through examination of photon noise limits and the absence of non-physical results, particularly in low-signal regimes such as the planet's night side. Furthermore, we assess the effectiveness of our systematic removal by analyzing the power spectral density (PSD) of the residuals. In an ideal scenario, the PSD should exhibit a flat, white-noise dominated spectrum at high frequencies, indicating successful removal of time-correlated noise. Any remaining structure in the PSD at lower frequencies may suggest the presence of unmodeled systematics or real astrophysical signals. This method effectively handles a wide range of observing modes and target brightnesses, maintaining model integrity across diverse datasets.

\subsection{Transit Model} 
The astrophysical model for the occultation of an exoplanet in front of its host star is based on the analytic expressions of \citet{Mandel2002}. For a transit, the parameters $R_{p}/R_{s}$, inclination, $i$, and $T_{mid}$ are left free during the analysis. Inclination is allowed to vary to help constrain the transit and eclipse durations. We focus on planet-specific free parameters, excluding stellar-dependent quantities like $a/R_{s}$ which rely on less certain stellar properties. The other transit parameters are fixed to values from the NASA Exoplanet Archive and are reported in the Excalibur data product RunID.PlanetName.system.finalize.parameters. Multiple sets of system parameters are generally available in the archive; for this study, we use the archive default selection. Where possible, missing quantities are derived from other parameters (e.g., $log(g)$ from planet mass and radius). Remaining missing fields are filled with non-default values, giving priority to more recent publications.

As the planet transits in front of the host star, brightness contrasts between the stellar limb and center regions, resulting in a predictable modulation of the shape of the transit. A quadratic limb darkening model is used and computed from the LDTK Toolkit in Python using the stellar parameters (see Figure~\ref{planet_parameters}). All planet parameters used in this analysis are available in the EXCALIBUR data product RunID.PlanetName.system.finalize.parameters, available on the IPAC EXCALIBUR portal. 

\begin{figure*}[t]
\includegraphics[scale=0.45]{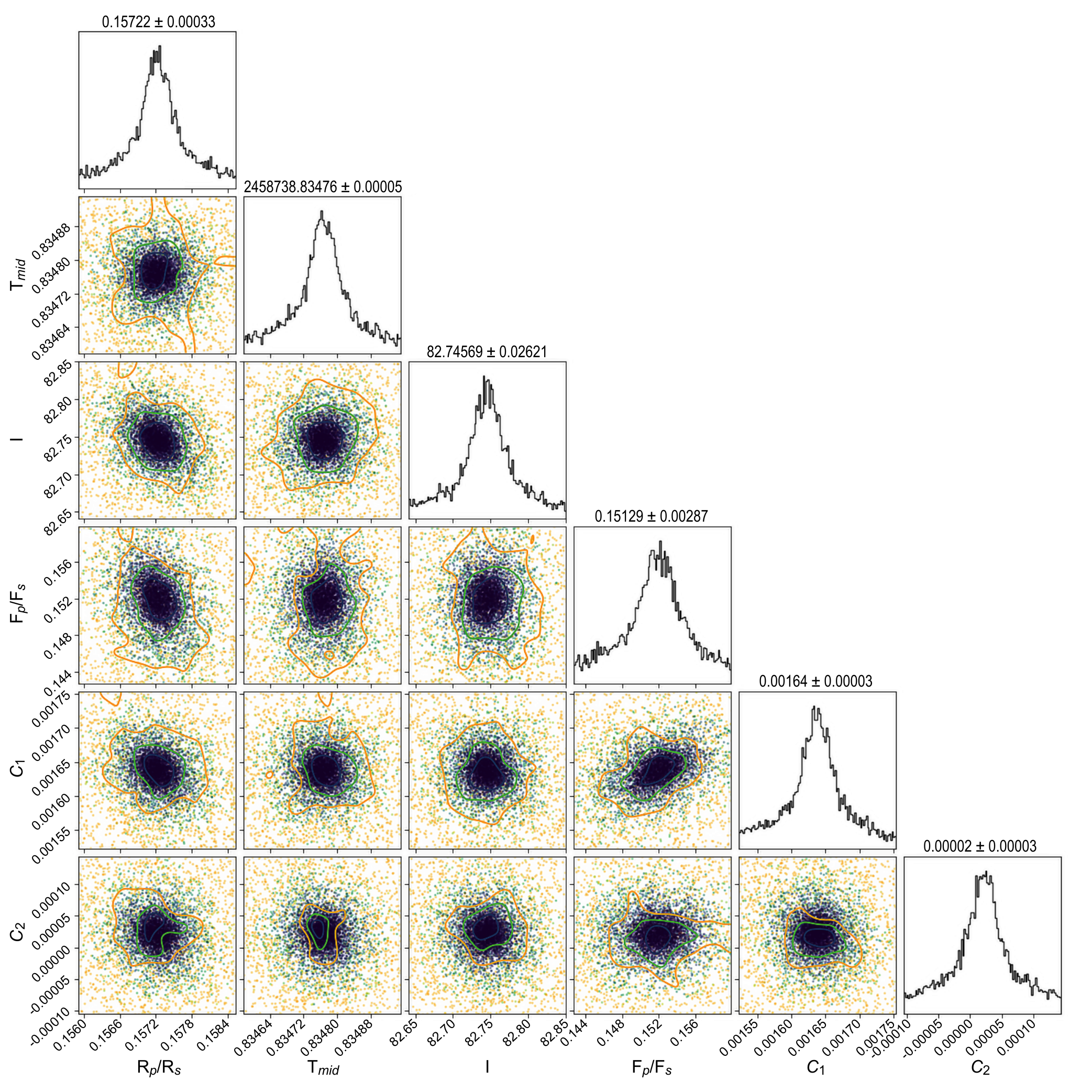}
\hspace{-1in}
\caption{A posterior distribution for a retrieval of the 4.5-$\mu$m phase curve parameters for WASP-43 b shown in Figure~\ref{36vs45}. The data points are color coded to the likelihood values with darker colors indicating higher likelihoods. The contours represent different sigma levels (orange = $>$3, green 2--3, and blue $<$1)  where the median and standard deviation of the distribution are reported in the title of each diagonal plot.}
\label{spitzer_posteriors}
\end{figure*}

\subsection{Eclipse Model}
As a planet passes behind its host star, the apparent change in brightness will be proportional to the planet's luminosity along with any reflected light from its atmosphere and/or surface. For the purposes of our analysis we assume the portion of reflected light between 3--5 $\mu$m is negligible and proceed to interpret brightness variations using a black-body radiation model. The eclipse depth is then proportional to the luminosity ratio between the planet and star which reduces to
\begin{equation} \label{eclipse_depth}
    Depth_{eclipse} = \frac{F_{p}}{F_{s}}  \frac{R_p^2}{R_s^2}
\end{equation}
and we specifically fit for $F_{p}/F_{s}$. If we were observing over all wavelengths, the flux term would converge to $\sigma T^{4}$, where $\sigma$ is the Stephan-Boltzmann constant and $T$ is the black-body temperature. However, channels 1 and 2 of Spitzer IRAC are each roughly 1 micron wide, therefore the flux is estimated by integrating a black-body function,  $B_\lambda$,  over the respective wavelength range like such:

\begin{equation}
\frac{F_p}{F_s}=\int_{\lambda}{\frac{B_\lambda(T_B)}{B_\lambda(T_{eff})} d\lambda}
\end{equation} A brightness temperature, T$_B$, for the planet can be derived after the fitting process once the flux ratio is measured with our eclipse model.

\subsection{Phase Curve Model}

The thermal phase curve of an exoplanet represents the variation in observed flux as a function of orbital phase, primarily due to the changing viewing geometry of the planet's day and night sides. We model these variations using a sinusoidal function:

\begin{equation}
C_0 + C_1 \, {\rm cos}\left(\frac{2\pi}{P}(t-t_{me})\right) + C_2 \, {\rm sin}\left(\frac{2\pi}{P}(t-t_{me})\right)
\end{equation}
where $C_0$ is an offset term normalized such that the relative flux during eclipse is 1, $P$ is the orbital period, $t$ is time, and $t_{me}$ is the time of mid-eclipse. The amplitude coefficients $C_1$ and $C_2$ represent the day-night brightness variation and hot-spot offset, respectively. This formulation is mathematically equivalent to the more traditional phase-offset representation:

\begin{equation}
C_0 + A \sin\left(\frac{2\pi}{P}t + \phi\right)
\end{equation}
where $A = \sqrt{C_1^2 + C_2^2}$ is the amplitude and $\phi = arctan(C_{2}/C_{1})$ is the phase offset. However, our mid-eclipse-referenced formulation provides several advantages. First, it allows direct physical interpretation of the day-night contrast ($C_1$) and hot-spot offset ($C_2$). Second, it enables straightforward implementation of physical constraints during the fitting process: we require $C_2 < C_1$ to ensure realistic hot-spot offsets, and $C_1$ must be less than the eclipse depth to guarantee positive night-side flux. While some studies have implemented additional harmonic terms to account for phase curve asymmetries, we adopt this simpler model to avoid over-fitting and to maintain clear physical interpretation of the fitted parameters. This model requires continuous observations spanning the full orbital period, which is best achieved through space-based observations.

\subsection{Nested Sampling}

When measuring subtle features like an exoplanet atmosphere, estimating an uncertainty on the measurement in a robust manner is just as important as the measurement itself. The retrieved parameters values and uncertainties (see Figure~\ref{spitzer_posteriors}) are derived from posterior distributions using a nested sampling algorithm. UltraNest is a Bayesian inference tool that uses the Monte Carlo strategy of nested sampling to calculate the Bayesian evidence allowing simultaneous parameter estimation and model selection \citep{Buchner2021}. A nested sampling algorithm is efficient at probing parameter spaces which could potentially contain multiple modes and pronounced degeneracies in high dimensions; a regime in which the convergence for traditional Markov Chain Monte Carlo (MCMC) techniques becomes incredibly slow (\citealt{Skilling2006}; \citealt{feroz2008}; \citealt{feroz2009}). The algorithm's efficiency is due to iteratively exploring smaller and smaller portions of a prior parameter space with sampling constraints being updated by local likelihood values (i.e., sample unlikely areas in a sparse manner and vice versa). The priors for our phase curve model have to be drawn from a conditional distribution in order to produce physical values. Given the analytic nature of our approach the model needs to reflect two conditions, (1) the stellar flux is normalized during eclipse and (2) the planet is contributing flux greater than or equal to 0 at all points in the orbit. In some cases, these constraints can lead to a cone-like structure in the posteriors and is particularly visible in the correlation plots with respect to $F_p/F_s$. 

\begin{figure*}[h]
\centering
\hspace{-0.25in} \\
\includegraphics[scale=0.29]{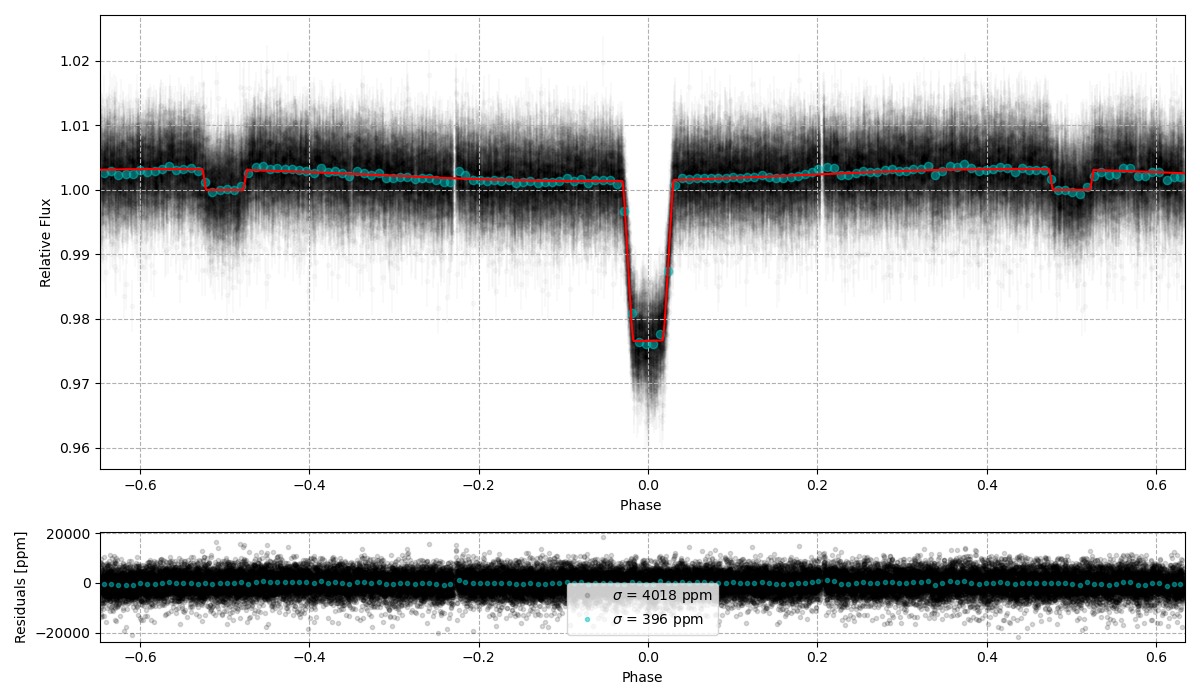}
\includegraphics[scale=0.29]{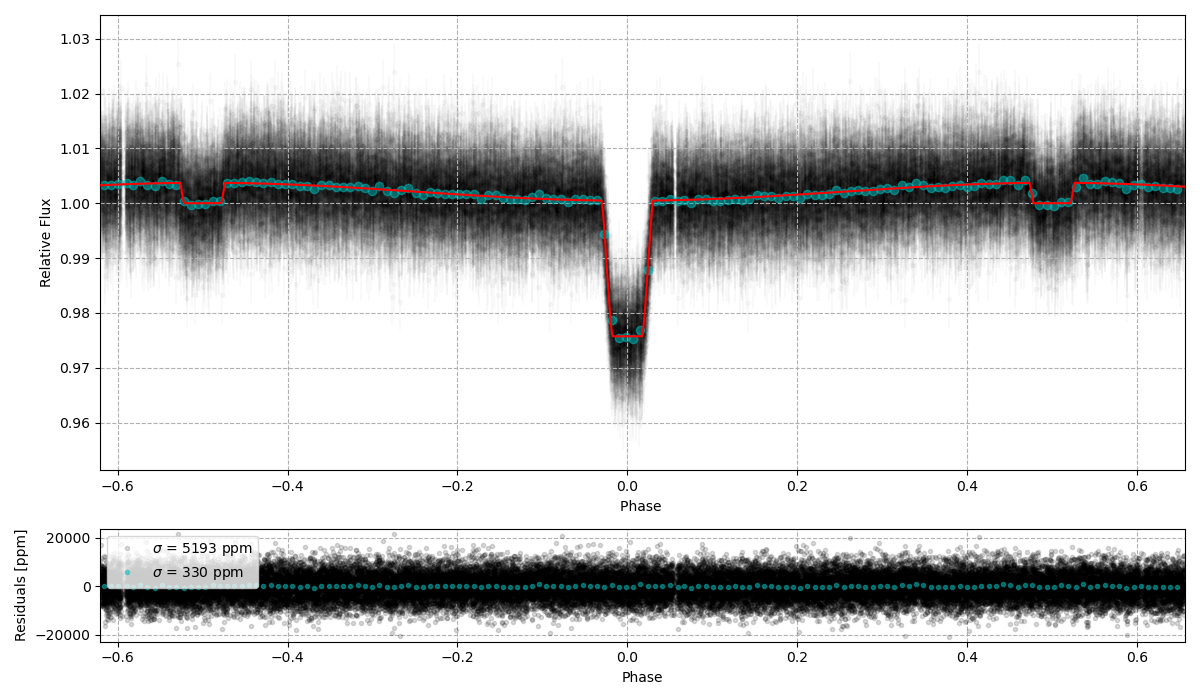}\\
\includegraphics[scale=0.32]{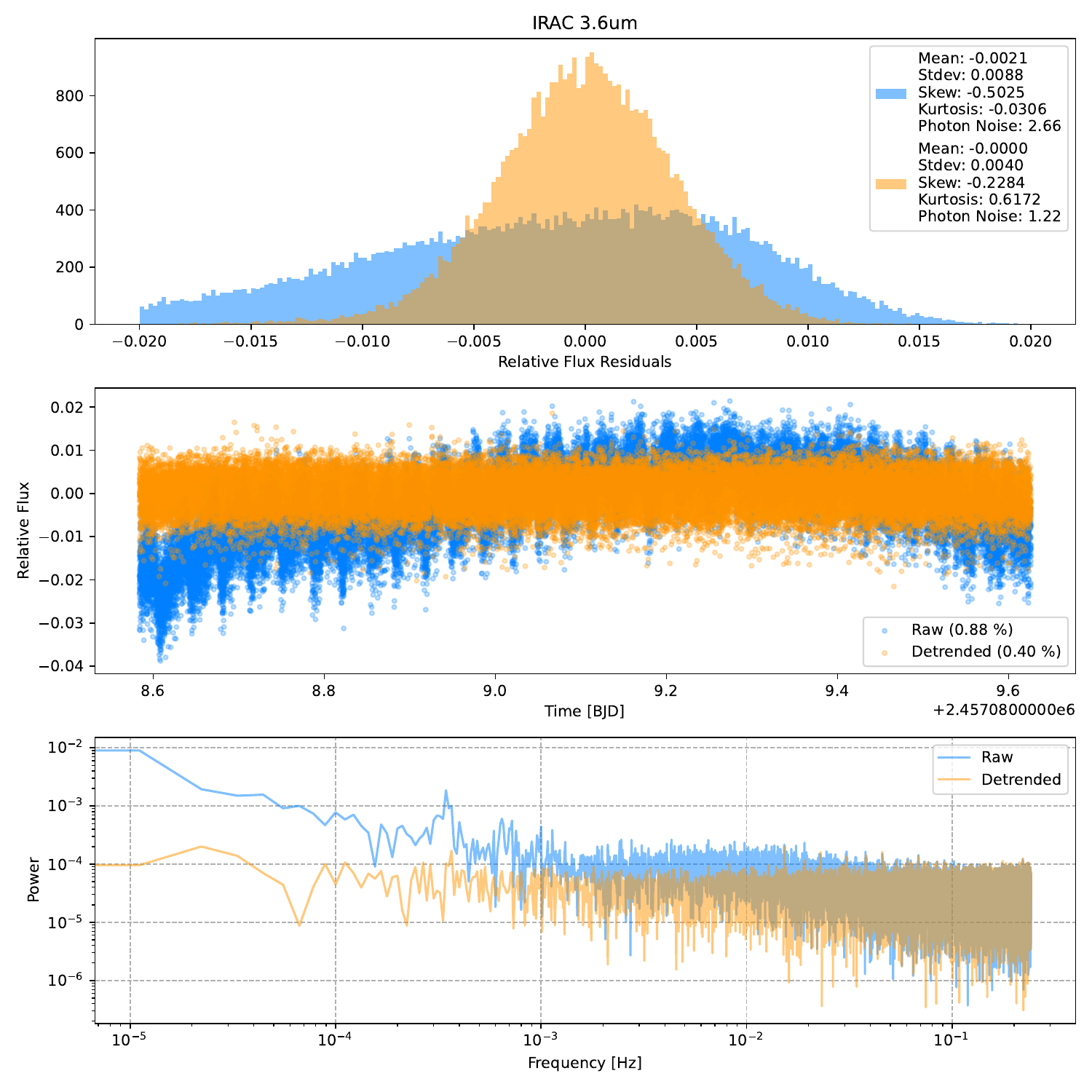}
\includegraphics[scale=0.32]{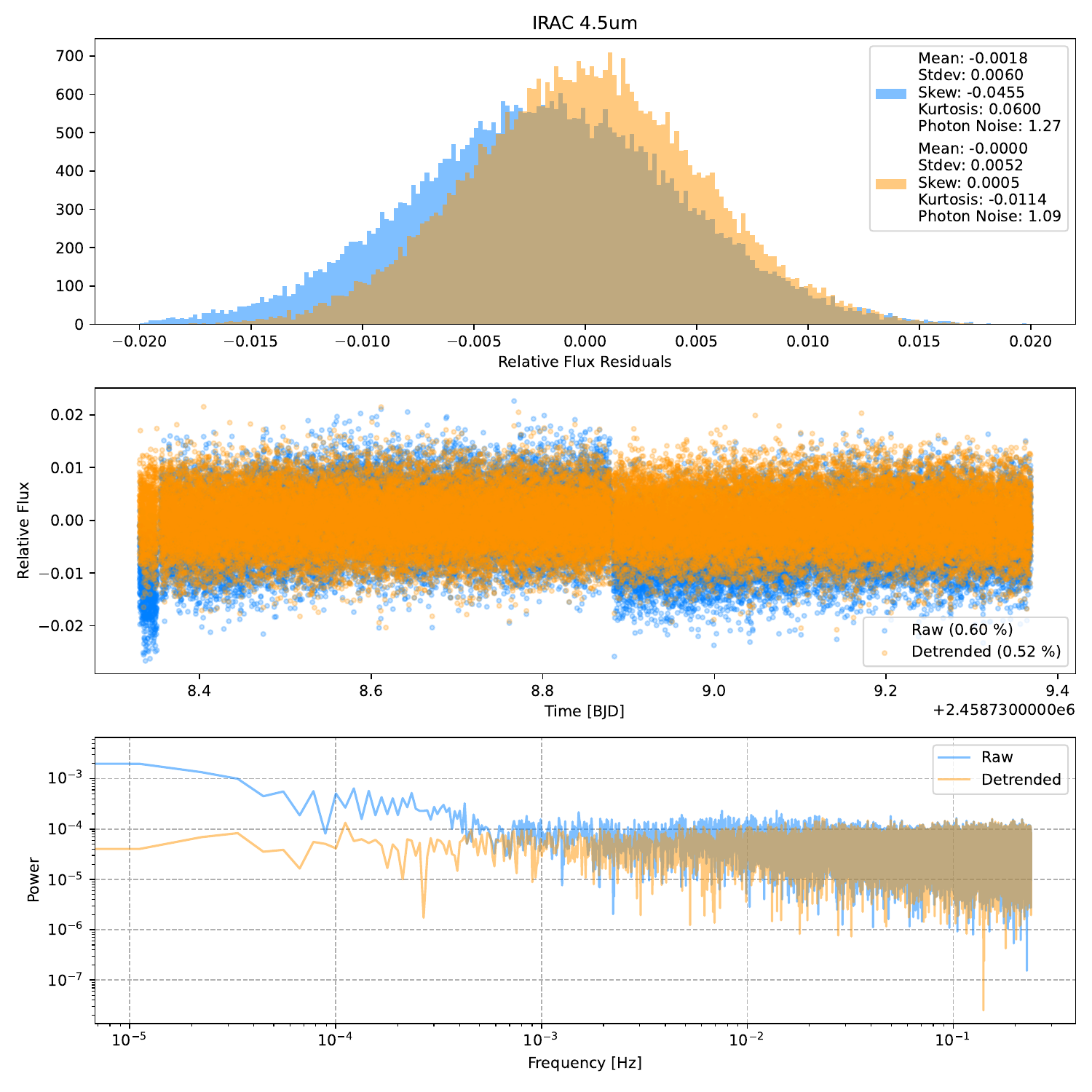}\\
\includegraphics[scale=0.35]{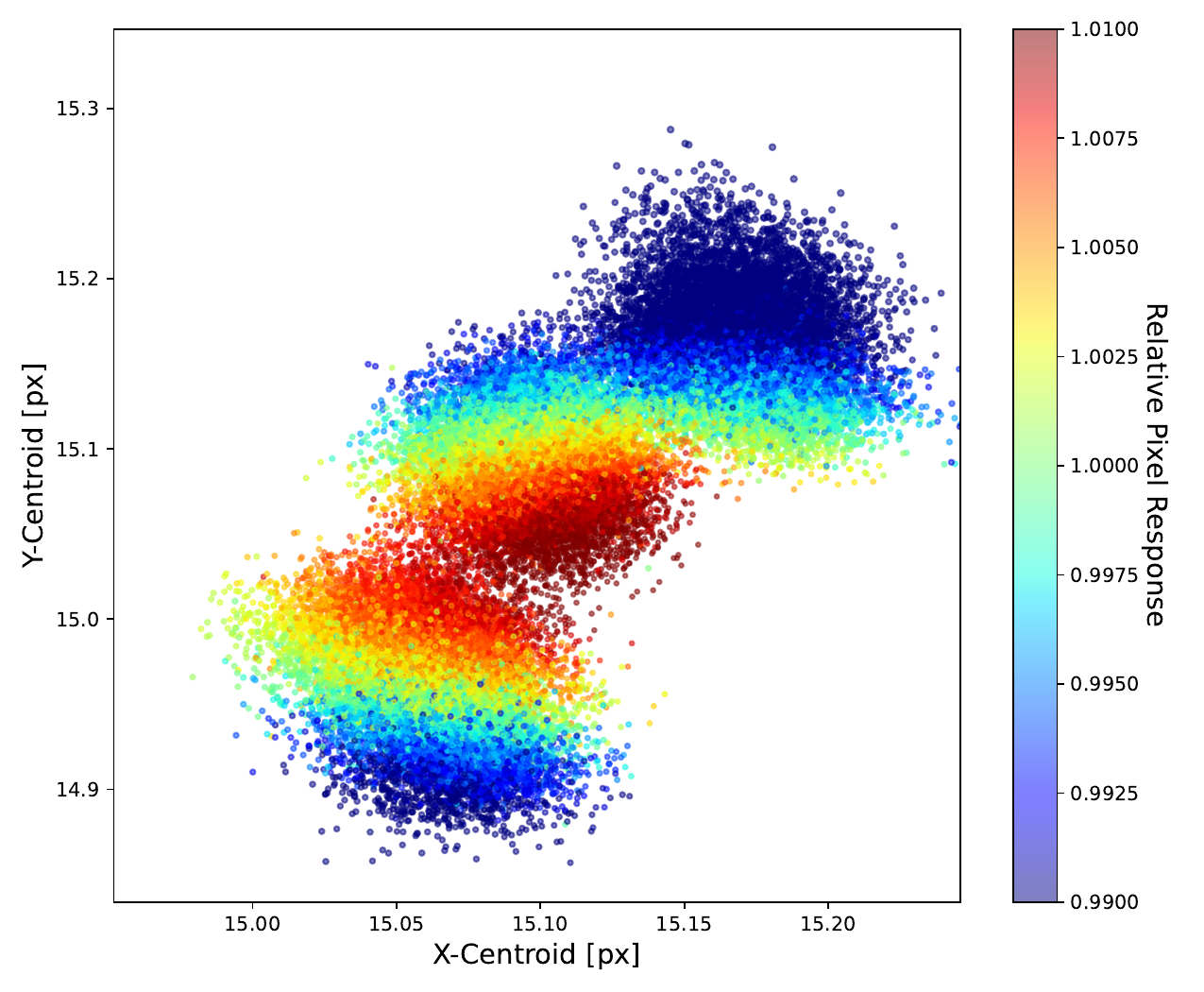}
\includegraphics[scale=0.35]{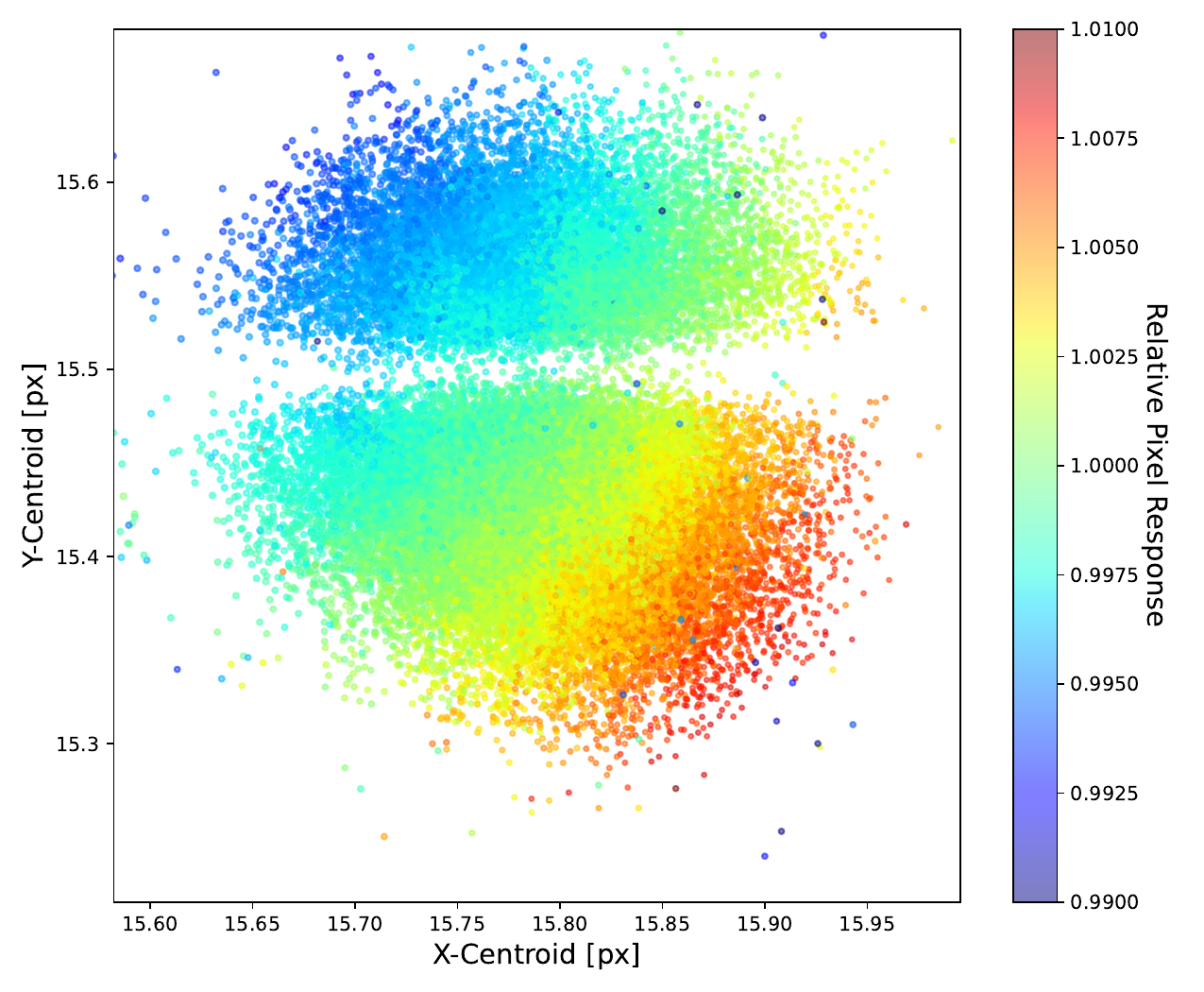}
\caption{WASP-43 b 3.6-$\mu$m (left) and 4.5-$\mu$m (right) phase curves and associated quality diagnostics.
Top row: same as in Figure~\ref{phasecurve} where the black data points are the observations at the native resolution after applying the pixel map to detrend for systematics. The colored data points are binned to a resolution of 5 minutes purely for visualization purposes. The red line indicates the best-fit transit model to the data.
Middle row: Quality diagnostics for each phase curve fit represented above it. The top panel compares a histogram of residuals after subtracting the best-fit model from the raw data (blue), whereas the orange data has been detrended with our instrument model before subtracting the astrophysical signal. The middle panel shows the timeseries of each residual. The bottom panel is the power spectral density between each of the residual timeseries. White noise dominated data will exhibit equal power at all frequencies. 
Bottom row: The map of the sub-pixel gain response pattern used for the instrument model. These plots are available as part of the associated data products for all the planets in this catalog.}
\label{36vs45}
\end{figure*}

\subsection{Instrument Model Performance Verification}

The combination of sufficient nearest neighbors and the conditional prior both helped in reducing over-fitting. None-the-less we monitor and quantify the quality of the instrument model by removing our determination of the astrophysical signature and then using three diagnostics contained in the Residual Statistics plots: (1) light curves before and after application of an instrument model, (2) histogram of residuals before and after application of an instrument model, and (3) the power spectral density (PSD) of the light curve before and after application of the instrument model (see Figure~\ref{36vs45}). The light curve diagnostic is mostly useful as a sanity check (orange band of points should be uniform and flat) and shows the flux modulation being absorbed by the instrument model (wiggles in the blue band). The histogram diagnostic shows the impact of the instrument model and quantifies the performance in terms of photon noise units. The PSD measures how well the instrument model has corrected systematic noise (slope and structure in the blue curve) to whiten the noise (orange curve trend should be flat for white noise), that is to correct the systematic noise at all frequencies present in the data. We consider systematic use and reporting of the pre- and post-instrument model PSDs an important innovation because it constitutes an explicit check that the instrument model has not created residuals at the low frequency portion of the PSD where the phase curve information is encoded.

The observations for targets in this catalog vary substantially in terms of level of systematics (as seen in the range of shapes of the blue data in the Residual Statistics diagnostic plots). These variations in systematics are accompanied by variations in where the image of the star was located on the detector -- which we visualize in maps of the centroid position and model for the relative pixel gain response (see Figure~\ref{36vs45}). We find our instrument modeling approach performs well in handling the range of systematic flux errors present in the catalog. We also find our approach works well on both the 3.6 and 4.5 $\mu$m channel data. By applying the Residual Statistics diagnostics across the catalog, we find that the instrument model produces residual light curves that are flat and visually free of artifacts,  Gaussian-like residuals typically around 1.2 times the photon noise, and PSDs that are flat. We do not see evidence of overfitting which would be indicated by sub-photon noise in the histogram diagnostic or by substantially reduced low frequency power in the PSD diagnostic. When we compare the 3.6 to the 4.5 $\mu$m data in detail, the 3.6~$\mu$m data generally has stronger systematic errors and instrument model performance is slightly worse, typically $\sim0.1$ photon noise. A direct comparison of the 3.6 and 4.5 $\mu$m data and Residual Statistics is shown in Figure~\ref{36vs45}.

To gain a sense of how our data reduction approach fits into the context of previous work, we  compare our phase curve parameter results for WASP-43 b to significant body of phase curve cross-comparison that has been done for this target \citep{stevenson2017,mendonca2018} (see Appendix). While our results are in family with previous work,  differences in methods, priors, and system parameter values, make it difficult to clearly isolate cause and effect. However, the analysis does underscore the need for a uniform data analysis approach in phase curve population studies. 

As another way to compare our results to previous work, we examined our results for data sets where previous authors have indicated problematic issues with the data reduction. An excellent example is the case of HD 149026 b, which was excluded from 4.5-$\mu$m phase curve study because of difficulties fitting the phase curve \citep{dang2024}. The phase curve for HD 1490926 b has been observed at both 3.6 and 4.5 $\mu$m. Our data reduction method was successful at modeling the instrument systematics for the 4.5-$\mu$m observations but failed on the more challenging 3.6-$\mu$m data (see Appendix Figure~\ref{fig:HD149026b}).

\section{Results}

We have 34 planets in our phase curve catalog, representing more than 125 days of Spitzer observing time. The catalog is summarized in Table~\ref{tab:phasecurve_catalog}. 
Of the 20 planets with both 3.6-$\mu$m and 4.5-$\mu$m phase curve measurements, two  have serious residual systematic errors in the 3.6-$\mu$m phase curve (HAT-P-23 b, HD 149026 b), one has incomplete phase curve measurements that do not adequately sample the eclipse (WASP-18 b), and one has a relatively low equilibrium temperature and shows no evidence of a secondary eclipse (GJ 1214 b). This leaves a subset of 16 planets with good phase curve measurements at 3.6 $\mu$m and 4.5 $\mu$m.

Since individual phase curves for each Spitzer visit have utility for a specific target (see top panels of Figures~\ref{36vs45}), we are making our phase curve catalog available to the community. Some of these targets have many phase curve visits and the catalog contains a total of 92 fully processed phase curves, all of which are available on the IPAC EXCALIBUR portal. The planets in this catalog have a large range of radii and equilibrium temperatures and not all of them are capable of generating phase curves that are detectable with Spitzer. HD 213885 b and K2-141 b represent the extreme non-detection examples in the catalog, each with multiple phase curve observations, none of which, individually, appear to have detected anything, including the transit. Nonetheless, we include the data for both of these targets in the catalog as they represent useful null hypothesis test cases of the data processing methods. The catalog contains numerous planets with high quality 4.5-$\mu$m phase curve measurements, such as KELT-20~b (see Figure~\ref{phasecurve}), that do not have corresponding 3.6-$\mu$m phase curve observations but which may be of interest for other phase curve related investigations.

\begin{table}[t]
  \caption{{\bf EXCALIBUR 2024 Spitzer Phase Curve Catalog:} Bold font indicates planets in the 2 Color Phase curve (2CP) sample. For planets in which there are multiple phase curve visits, the visit number indicated in parenthesis is used for the $C_{1}$ and $C_{2}$ coefficient values.} 
  \label{tab:phasecurve_catalog}
  \begin{center}
\begin{tabular}{lcccc} \hline \hline
Planet   & Band & phase& Band & phase \\ 
         &  [$\mu$m] & curves & [$\mu$m] & curves \\     \hline 
      CoRoT-2 b  &     &       &  4.5  & 1     \\
      GJ 1132 b  &     &       &  4.5  & 2     \\
      GJ 1214 b  & 3.6 & 1     &  4.5  & 11    \\
{\bf HAT-P-2 b}  & 3.6 & 1     &  4.5  & 2 (1) \\
      HAT-P-23 b & 3.6 & 1     &  4.5  & 1     \\
{\bf HAT-P-7 b}  & 3.6 & 1     &  4.5  & 1     \\
      HD 149026 b& 3.6 & 1     &  4.5  & 1     \\
      HD 213885 b&     &       &  4.5  & 2     \\
      K2-141 b   &     &       &  4.5  & 10    \\
{\bf KELT-1 b}   & 3.6 & 1     &  4.5  & 1     \\
{\bf KELT-14 b}  & 3.6 & 1     &  4.5  & 1     \\
      KELT-16 b  &     &       &  4.5  & 1     \\
      KELT-20 b  &     &       &  4.5  & 1     \\
      KELT-7 b   &     &       &  4.5  & 1     \\
{\bf KELT-9 b}   & 3.6 & 1     &  4.5  & 1     \\
      LHS 3844   &     &       &  4.5  & 9     \\
      MASCARA-1 b&     &       &  4.5  & 1     \\
{\bf Qatar-1 b}  & 3.6 & 1     &  4.5  & 1     \\
{\bf Qatar-2 b}  & 3.6 & 1     &  4.5  & 1     \\
      TrES-3     &     &       &  4.5  & 1     \\
{\bf WASP-12 b}  & 3.6 & 2 (2) &  4.5  & 2 (2) \\
{\bf WASP-121 b} & 3.6 & 1     &  4.5  & 1     \\
{\bf WASP-14 b}  & 3.6 & 1     &  4.5  & 2 (1) \\
      WASP-140 b &     &       &  4.5  & 1     \\
      WASP-18 b  & 3.6 & 1     &  4.5  & 1     \\
{\bf WASP-19 b}  & 3.6 & 1     &  4.5  & 1     \\
{\bf WASP-33 b}  & 3.6 & 1     &  4.5  & 1     \\
      WASP-34 b  &     &       &  4.5  & 1     \\
{\bf WASP-43 b}  & 3.6 & 2 (1) &  4.5  & 3 (2) \\
{\bf WASP-52 b}  & 3.6 & 2 (2) &  4.5  & 1     \\
       WASP-74 b &     &       &  4.5  & 1     \\
{\bf WASP-76 b}  & 3.6 & 2 (1) &  4.5  & 1     \\
{\bf WASP-77 b}  & 3.6 & 1     &  4.5  & 1     \\ 
       WASP-95 b &     &       &  4.5  & 1     \\ \hline
\end{tabular}
\\ 
\end{center}
\end{table}

\section{Discussion}

The parameters $R_{p}/R_{s}$, $F_{p}/F_{s}$, $C_{1}$, and $C_{2}$ (see Equations 4, 5, and 6) are the parameters that determine the observed properties of the phase curve. In our study, there are 16 planets that have good quality phase curve measurements at both 3.6 and 4.5 $\mu$m  (see discussion in Section 4) and we term this group the Spitzer two-color phase curve (2CP) sample (shown in bold in Table~\ref{tab:phasecurve_catalog}). In cases where there were multiple observations of a phase curve, we selected the phase curves (identified in Table~\ref{tab:phasecurve_catalog}) that had the lowest level of residual systematics in the low frequency portion of the power spectral density diagnostic as those are the frequencies that couple to the phase curve amplitude and phase. The 2CP sample phase curve amplitude, $A$ (measured in ppm), and offset, $\phi$ (measured in degrees), are listed in Table~\ref{tab:phasecurve_amp_phase} in the Appendix.

Our expectation is that phase curve properties should depend on physical parameters such as equilibrium temperature, $T_{eq}$, orbital period, $P$, surface gravity $log(g)$, and planet radius, $R_{p}$, and these values are automatically scraped by our pipeline from the NASA Exoplanet Archive and the references noted as part of the system.finalize.parameter data product (see Figure~\ref{planet_parameters}). However, in the course of preparing this manuscript, we found that $T_{eq}$ values reported by the NASA Exoplanet Archive are not formulated in a consistent way, including for planets in the 2CP sample. Therefore, we calculated the values for $T_{eq}$ using the reported stellar temperature, stellar radius, and planet semi-major axis values reported in the NASA Exoplanet Archive (all of these parameter values and the associated references are listed in the system.finalize.parameters data product). The 2CP sample values for $T_{eq}$, $P$, and $R_{p}$ are shown in Table~\ref{tab:phasecurve_planet_proterties} in the Appendix. Unsurprisingly, the planets in this sample are in the hot-Jupiter, or ultra hot-Jupiter, categories. The sum of the orbital periods for these planets is 27.2 days and thus, with overheads, represents approximately two months of Spitzer telescope time to acquire this unique data set. 

\begin{figure}[h]
    \centering
    \includegraphics[width=0.45\textwidth]{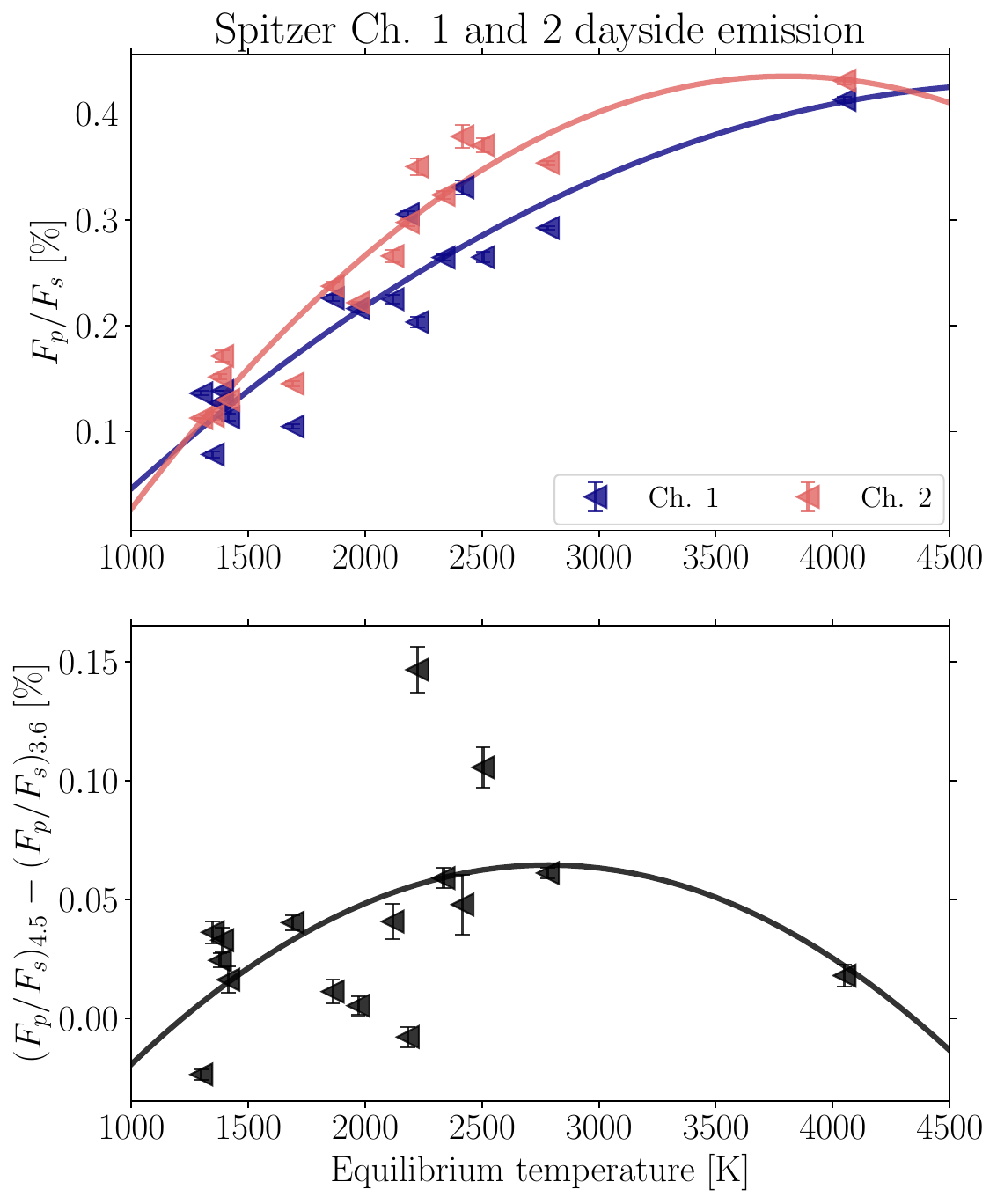}
    \caption{The sample of 16 planets with robust 3.6- and 4.5-$\mu$m phase curves shows the expected trend of increasing $F_{p}/F_{s}$ with increasing equilibrium temperature, though with evidence for non-linearity in the dependence of $F_p/F_s$ with $T_\mathrm{eq}$ (top). For the planets in this sample, the 4.5 $\mu$m $F_{p}/F_{s}$ values are almost always larger than the 3.6 $\mu$m $F_{p}/F_{s}$ values (bottom).}
    \label{fig:Fp/Fs}
\end{figure}

We begin our investigation of the 2CP sample with an examination of $F_{p}/F_{s}$ dependence on temperature. As expected, there is a well-defined trend for the dependence of $F_p/F_s$ on $T_\mathrm{eq}$. However, this trend is measurably different for the 3.6-$\mu$m and 4.5-$\mu$m flux values, with a 3.6-$\mu$m $F_{p}/F_{s}$ flux deficit relative to the 4.5-$\mu$m $F_{p}/F_{s}$ values growing as a function of temperature and then decreasing for KELT-9b (see Figure~\ref{fig:Fp/Fs}). Note that KELT-9b has published TESS and Spitzer phase curves that both show this relatively low dayside flux and higher heat redistribution compared to an extrapolation of cooler gas giants \citep{Wong:2019aa,Mansfield:2020aa}. However, further phase curve observations of ultra-hot Jupiters between KELT-9b (e.g., TOI-2109b, \citet{Wong:2021td}) and the bulk of the ultra-hot Jupiter population are required to confirm this downturn. This behavior was previously reported in an analysis of Spitzer exoplanet eclipse measurements \citep{baxter2020} and we identify the same trend with this sample of planets with well-characterized phase curves. The relative increase in 4.5-$\mu$m $F_{p}/F_{s}$ with respect to 3.6-$\mu$m $F_{p}/F_{s}$ as a function of $T_{eq}$ was interpreted as due to CO emission in the presence of a thermal inversions \citep{baxter2020}. The CO emission scenario implies that the 4.5-$\mu$m band is probing higher in the exoplanet atmosphere (i.e., lower pressures and higher temperatures in the presence of an inversion) than the 3.6-$\mu$m band.

For analysis of the phase curve properties, we examined the $A$, $\phi$, $\Delta A$ and $\Delta \phi$ parameters ($\Delta A = A^{4.5 \mu m}-A^{3.6 \mu m}$  and $\Delta \phi = \phi^{4.5 \mu m} - \phi^{3.6 \mu m}$ for possible correlations with physical parameters $T_{eq}$, $P$, $R_{p}$, $log(g)$, and various combinations of these physical parameters which we term ``combination parameters.'' When discussing $A$ and $\phi$ values for specific Spitzer wavelength bands, we note the IRAC channel as a superscript (e.g., $A^{4.5\mu m}$). For the correlation search, we uniformly applied a first order polynomial fit and used the R$^{2}$ statistic to measure the strength of the correlation -- see Table~\ref{tab:Rsquared} for a summary of results.

\begin{table*}[t]
  \caption{R$^{2}$ Values for Physical and Combination Parameters; maximum significant correlations identified in bold.}
  \label{tab:Rsquared}
  \begin{center}
\begin{tabular}{l|ccc|ccc} \hline \hline
Parameter & 
$A^{3.6\mu m}$ & $A^{4.5\mu m}$ & $\Delta A$ &
$\phi^{3.6\mu m}$ & $\phi^{4.5\mu m}$ & $\Delta\phi$\\ 
\hline
$T_{eq}$ & 0.005 & 0.036 & 0.012 & \bf 0.286 & 0.062 & \bf 0.163 \\
$P$ & 0.221 & \bf 0.350 & 0.001 & 0.205 & 0.081 & 0.083 \\
$R_p$ & 0.103 & 0.084 & 0.010 & 0.012 & 0.048 & 0.001 \\
$log(g)$ & \bf 0.331 & 0.169 & 0.087 & 0.032 & \bf 0.146 & 0.006 \\
\hline
$R_{p}P$ & 0.155 & 0.270 & 0.003 & 0.168 & 0.043 & 0.088 \\
$R_{p}PT_{eq}$ & 0.079 & 0.083 & 0.002 & 0.007 & 0.001 & 0.005 \\
$T_{eq}/P$ & 0.152 & 0.330 & 0.014 & 0.380 & 0.046 & 0.267 \\
$T^{2}_{eq}/(R_{p}P)$ & 0.027 & 0.117 & 0.025 & \bf 0.427 & 0.056 & \bf 0.292 \\
$T^{1/2}_{eq}$ & 0.012 & 0.051 & 0.011 & 0.252 & 0.051 & 0.148 \\
$T^{2}_{eq}$ & 0.000 & 0.011 & 0.016 & 0.343 & 0.081 & 0.187 \\
$T^{5/2}_{eq}$ & 0.002 & 0.004 & 0.017 & 0.361 & 0.088 & 0.195 \\
$T^{4}_{eq}$ & 0.019 & 0.001 & 0.020 & 0.383 & 0.098 & 0.201 \\
($T^{2}_{eq} log(g))/(R_{p}P)$ & 0.002 & 0.046 & 0.029 & 0.309 & 0.031 & 0.228 \\
$R_{p}PT_{eq} \times log(g)$ & 0.182 & 0.177 & 0.009 & 0.000 & 0.007 & 0.003 \\
\hline
$R_p^{0.5} T_{eq}^{-0.1} g^{-0.5} P^{-1}$ & \bf 0.596 & 0.405 & 0.092 & 0.044 & 0.087 & 0.000 \\
$R_p^{1.1} T_{eq}^{-0.4} g^{0} P^{-0.9}$ & 0.426 & \bf 0.589 & 0.000 & 0.089 & 0.019 & 0.050 \\
\hline
\end{tabular}
\end{center}
\label{tab:Rsquared}
\end{table*}

The strongest correlation between phase curve amplitude and individual physical parameters ($T_\mathrm{eq}$, $P$, $R_{p}$, $log(g)$) is different for 3.6 $\mu$m and 4.5 $\mu$m; $A^{3.6 \mu m}$ is most strongly correlated with $log(g)$ and $A^{4.5 \mu m}$ is most strongly correlated with $P$.  Similarly, $\phi^{3.6 \mu m}$ and $\phi^{4.5 \mu m}$ correlate most strongly with different physical parameters. This suggests that there may be differences in the underlying physical mechanism controlling the temperature of the emitting regions at 3.6 and 4.5 $\mu$m. We find the phase offset at 4.5 $\mu$m correlates most strongly with $log(g)$, which is in agreement with previous work, based on a smaller sample, reporting a tentative trend for 4.5-$\mu$m phase curve offset and $log(g)$ \citep{may2022}. We find $T_\mathrm{eq}$ and $log(g)$ are the most important physical parameters for determining the phase curve offset at 3.6 $\mu$m and 4.5 $\mu$m respectively. We find some degree of correlation between orbital period and phase offset at 3.6-$\mu$m but not at 4.5-$\mu$m as reported by \cite{may2022}.
 
Previous work considering an atmospheric composition and range of temperature profiles relevant for a hot-Jupiter found that the 3.6-$\mu$m contribution function peaks at systematically higher pressures than the 4.5-$\mu$m contribution function \citep{swain2009b,Dobbs-Dixon:2017aa,stevenson2017}; this is consistent with the interpretation given to the chromatic dependence of $F_{p}/F_{s}$ \citep{baxter2020}. We conclude that $log(g)$ is an important determinative physical variable for phase curve amplitude at higher pressures, whereas at lower pressures, $log(g)$ plays no significant role in establishing phase curve amplitude. This could imply that the 3.6~$\mu$m phase curve properties are influenced by the presence of a cloud deck, but further work would be needed to develop evidence to support this idea.

\begin{figure}[h]
    \centering
    \includegraphics[width=0.5\textwidth]{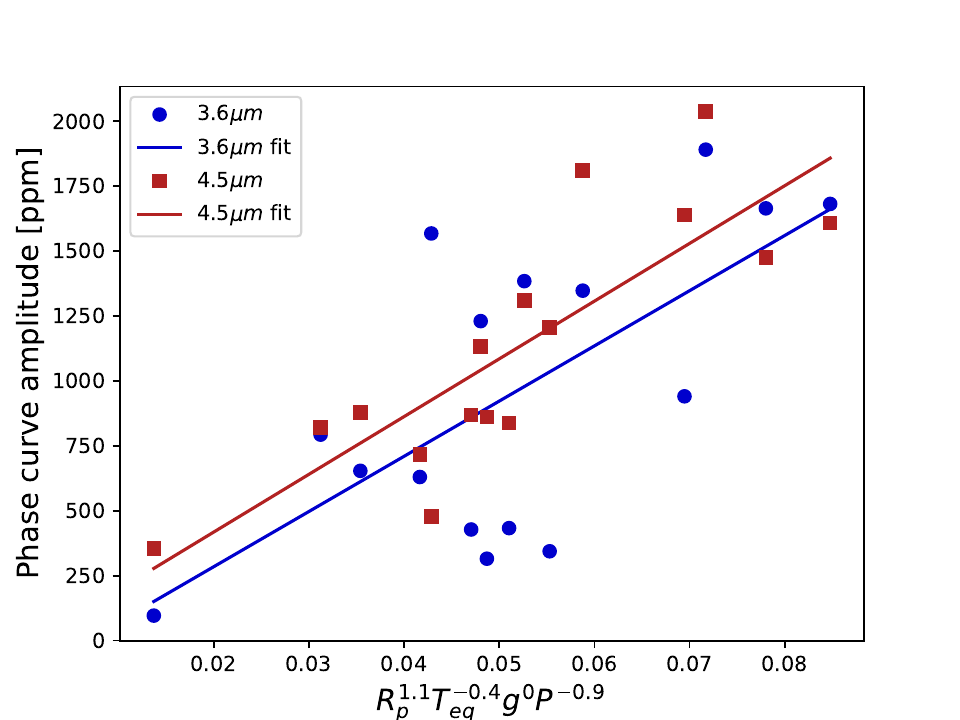}
    \caption{The strongest correlation we identified for the 4.5-$\mu$m phase curve amplitudes ($R^{2}=0.589$, red squares) is with the combination of physical parameters, suggesting that multiple physical parameter participate in establishing thermal phase curve properties. Also plotted for comparison are the 3.6-$\mu$m phase curve amplitudes ($R^{2}=0.426$, blue circles).}
    \label{fig:c1_T2RP}
\end{figure}

We find that combinations of physical parameters can increase the strength of the correlation, above the single physical parameter correlation value, for $A^{4.5 \mu m}$, $\phi^{3.6 \mu m}$, $\Delta \phi$, and  parameters. This finding suggests that multiple physical parameters have a role in establishing phase curve properties. In addition to trying a variety of combination parameters, we used the Gauss-Markov (GM) method to identify the optimal combination parameters having the form $R_{p}^{w} T_{eq}^{x} g^{y} P^{z}$ for $A^{3.6 \mu m}$ and $A^{4.5 \mu m}$ respectively. The GM-identified combination parameters are listed in the last two rows of Table~\ref{tab:Rsquared}. The correlation search shows that for both single parameter and combination parameters, $A^{3.6 \mu m}$ and $A^{4.5 \mu m}$ have a different dependence on the physical parameters. In particular, the $A^{4.5 \mu m}$ does not seem strongly influenced by the $log(g)$ parameter whereas $log(g)$ is the single most important physical parameter for $A^{3.6 \mu m}$ in our analysis. To visualize the correlations, we plot the GM-identified combination parameter $R_{p}^{1.1} T_{eq}^{-0.4} g^{0} P^{-0.9}$ (see Figure~\ref{fig:c1_T2RP}); while the linear trend is clear, the scatter about the trend suggests that there could be additional physical variables that are not captured in this combination parameter. Of the GM-identified combination parameters that have the best correlations with $A^{3.6 \mu m}$ and $A^{4.5 \mu m}$ respectively, both have exponents for P that have a larger absolute value than the exponents for T; this implies that rotation period is more important than temperature in establishing the phase curve amplitude, leading to the stronger dependence on P in the GM search.

Recent GCM studies \citep{roth2024} predict that the phase curve offset should be a strong function of $T_\mathrm{eq}, P, log(g)$, and metallicity. We find observationally that the physical parameter most correlated with phase curve offset is different at 3.6 $\mu$m and 4.5 $\mu$m. As with amplitude, the fact that the individual physical parameter with the strongest correlation is different for the 3.6-$\mu$m and 4.5-$\mu$m phase offset, suggests that different physical processes are driving the temperatures of the 3.6-$\mu$m and 4.5-$\mu$m emitting regions. The observed phase curve correlations imply that more than one physical parameter is important for determining phase curve offset and that the role of $P$ is sub-dominant. The correlation results also indicate that the phase curve offset is not controlled by a simple longitudinal dependence of the vertical temperature profile (because the $\phi$ parameter should be similarly affected at 3.6 $\mu$m and 4.5 $\mu$m and thus the $\Delta \phi$ parameter should result in decorrelation). This implies that some other factor, such as Doppler-shifting of the planetary-scale Matsuno-Gill pattern \citep{Hammond:2018aa,Lewis:2022aa} or longitudinally dependent opacity structure due to compositional changes (e.g., inhomogeneous clouds, \citealp{Parmentier:2021tt}) may be the primary influence in determining the phase curve offset.

\section{Conclusion}

The Infrared Array Camera on the Spitzer Space Telescope observed exoplanets for over a decade and has a large archive of observations. We present a catalog of exoplanet phase curves analyzed in a uniform way using a version of the pixel-map technique based on nearest neighbors regression with a Gaussian kernel. We report phase curves for a catalog of 34 planets, some of which have multiple phase curve measurements. Within the catalog, 16 planets have high-quality, full orbit phase curve measurements in both the 3.6-$\mu$m and 4.5-$\mu$m IRAC filters. We use the phase curve model parameter, $F_{p}/F_{s}$, and the phase curve amplitude and phase, constructed from the $C_{1}$ and $C_{2}$ phase curve harmonic parameters, to search for trends and correlations. 

We consider four physical parameters, $T_{eq}$, $P$, $R_{p}$, and $log(g)$, which are candidates for influencing properties of exoplanet thermal phase curves. Using the unique set of 16 planets that have high-quality 3.6-$\mu$m and 4.5-$\mu$m phase curves, we show that the correlations of the phase curve amplitude and phase with physical parameters is fundamentally different for the 3.6-$\mu$m and 4.5-$\mu$m bands. The 4.5-$\mu$m phase curve amplitude is most strongly correlated with the planetary orbital period where as the 3.6-$\mu$m phase curve amplitude is most strongly correlated with the planetary $log(g)$. We also show that combinations of physical parameters can, in some cases, exhibit a greater degree of correlation than individual physical parameters. Taken together, this suggests that different combinations of physical parameters, probably corresponding to different physical mechanisms, are responsible for establishing the phase curve properties at 3.6 $\mu$m and 4.5 $\mu$m. These mechanisms are multi-faceted, likely including the competition between wave adjustment and radiative cooling \citep{Perez-Becker:2013fv,Parmentier:2021tt}, Lorentz forces or other sources of frictional drag \citep{Rogers:2020,Koll:2017,Beltz:2022aa}, thermal inversions \citep{Roth:2024aa}, molecular dissociation \citep{Bell:2018aa,Roth:2021un}, and cloud coverage \citep{Keating:2018aa,Beatty19,Gao:2021vp,Roman:2021wl}. Our results suggest that the generation of spectral thermal phase curves for powerfully heated gas giant exoplanets is a complex phenomenon in which different physical processes may dominate at different planetary regimes and atmospheric pressures.

\section*{Acknowledgements}
The research was carried out at the Jet Propulsion Laboratory, California Institute of Technology, under a contract with the National Aeronautics and Space Administration (80NM0018D0004).

\appendix

Here we provide additional information relating to the EXCALIBUR 2024 phase curve catalog.

\section{Target information}

\begin{table}[h]
  \caption{EXCALIBUR 2024 Spitzer Phase Curve Catalog with the associated Spitzer program ID, proposal PI, and relevant publication.} 
  \label{tab:phasecurve_catalog_pid}
  \begin{center}
\begin{tabular}{lccc} \hline \hline
Planet   & Program ID & Program PI & Citation \\   \hline 
      CoRoT-2 b  & 11073 & Cowan & \cite{dang2018} \\
      GJ 1132 b  & 12082 & Dittman &  \cite{dittmann2017} \\
      GJ 1214 b  & 14130, 14233, 14253 & Stefanon, Morishita &  -- \\
      HAT-P-2 b  & 60021 & Knutson & \cite{Lewis2013} \\
      HAT-P-23 b & 13038 & Stevenson & \cite{dang2024} \\
      HAT-P-7 b  & 60021 & Knutson & \cite{wong2016} \\
      HD 149026 b& 60021 & Knutson & \cite{zhang2018} \\
      HD 213885 b& 14084 & Crossfield & -- \\
      K2-141 b   & 14135 & Kreidberg & \cite{zieba2022} \\
      KELT-1 b   & 11095 & Beatty &  \cite{beatty2019}  \\
      KELT-14 b  & 13038 & Stevenson & \cite{dang2024} \\
      KELT-16 b  & 14059 & Bean & \cite{bell2021}    \\
      KELT-20 b  & 14059 & Bean & \cite{dang2024}  \\
      KELT-7 b   & 14059 & Bean & --  \\
      KELT-9 b   & 14059 & Bean & \cite{Mansfield:2020aa}  \\
      LHS 3844 b & 14204 & Kreidberg  & \cite{kreidberg2019}   \\
      MASCARA-1 b& 14059 & Bean & \cite{bell2021}     \\
      Qatar-1 b  & 13038 & Stevenson & \cite{keating2020}  \\
      Qatar-2 b  & 13038 & Stevenson & \cite{may2022}  \\
      TrES-3 b   & 14059 & Bean & \cite{dang2024}   \\ 
      WASP-12 b  & 70060,90186 & Machalek, Todorov & \cite{cowan2012,Bell2019} \\
      WASP-121 b & 13242 & Evans & \cite{morello2023}  \\
      WASP-14 b  & 80073 & Knutson & \cite{wong2015} \\
      WASP-140 b & 14059 & Bean & \cite{may2022}   \\
      WASP-18 b  & 60185 & Maxted & \cite{maxted2013}  \\
      WASP-19 b  & 80073 & Knutson & \cite{wong2016} \\
      WASP-33 b  & 80073 & Knutson & \cite{zhang2018}  \\
      WASP-34 b  & 14059 & Bean & \cite{may2022}  \\
      WASP-43 b  & 11001 & Stevenson & \cite{stevenson2017,may2020} \\
      WASP-52 b  & 13038 & Stevenson & \cite{may2022}  \\
      WASP-74 b  & 14059 & Bean & \cite{dang2024}  \\
      WASP-76 b  & 13038 & Stevenson & \cite{may2021}    \\
      WASP-77 b  & 13038 & Stevenson & \cite{dang2024}   \\ 
      WASP-95 b  & 14059 & Bean & --  \\ \hline
\end{tabular}
\\ 
\end{center}
\end{table}

\newpage

\section{WASP 43 \lowercase{b} Results Comparison}

A comparison of phase curve results obtained using different instrument models also needs to consider a comparison of input parameters because the instrument model is designed to absorb differences between the astrophysical model and the measurements. Ideally, systems parameters should be fully self consistent so that the orbital solution is consistent with Kepler's third law, although this is not possible in all cases. Because of the potential for specific values of system parameters to impact the results, our approach is to carefully document the parameters we use in our analysis. WASP-43 b has been used extensively by the community as a Spitzer phase curve target for comparing different data reduction methods \citep{stevenson2017,mendonca2018,morello2019,may2020,bell2021,dang2024} and we continue this tradition. When comparing the different published results for WASP-43 b, what immediately stands out is the lack of information about the systems parameters used in the astrophysical model. In our survey of the aforementioned six Spitzer phase curve publications for WASP-43 b, only two of them lists the parameters used \citep{morello2019,dang2024}, and only two of them describe a selection method for the astrophysical parameters \citep{bell2021,dang2024}. The remainder of the papers do not discuss the parameters used in the light curve modeling; that is there is no list for the parameter values, no citation of the parameter values, and no discussion of parameter value selection criteria. The two most recent papers use the ``most precise'' values for each parameter \citep{bell2021,dang2024}; in our view, this approach is undesirable because it can lead to orbital parameters that are fundamentally inconsistent, even when a consistent set is available; this potential hazard is demonstrated in WASP-43 b where a self-consistent set of system parameters is available \citep{hellier2011} but choosing the most precise values potentially creates inconsistencies.

To compare our results for WASP-43 b to previous results, we use the summary from \cite{Bell2019} which compares the phase curve parameter values $R_{p}/R_{s}$, $F_{p}/F_{s}$, $A$, and $\phi$. In our assessment, the lack of information about the light curve modeling parameters prevents detailed inference about the relative merit of different methods based on comparing the results. We also note that in addition to differences in system parameters and methods, at least one group applies a 3000 pppm prior to the phase curve amplitude \citep{bell2021}. The combination of different methods, different system parameters, and different priors makes detailed comparison of the results difficult; a single factor, such as the method, can not be unambiguously associated with a change in a resulting parameter, such as the value of $F_{p}/F_{s}$. In comparing our results with previous work, we do find that our phase curve parameter values for WASP-43 b overlap previous results for $R_{p}/R_{s}$, $F_{p}/F_{s}$, and $A$, while our value for $\phi$ is smaller than previous results. This exercise of comparing our phase curve parameter results with previously published work highlights the need for statistical studies to take a uniform analysis approach in which the same method is applied to all targets in the sample.

\begin{sidewaystable}[h]
  \caption{Comparison of WASP-43 b 4.5 $\mu$m Phase Curve Results}
\label{tab:W43_compare}    
\begin{center}
\begin{tabular}{cccccccc} \hline \hline
 parameter & this work                  & Dang24                   & Bell21 & May21 & Morello19 & Mendonca18 & Stevenson17 \\   \hline 
Stellar radius -  $R_{s}$ & $0.6^{0.03}_{-0.04}$      [1]& $0.6^{0.03}_{-0.04}$   [1]&?&?& $0.6^{0.03}_{-0.04}$ [1]&?&? \\
Steller mass - $M_{s}$      & $0.58^{0.05}_{-0.58}$     [1]& not used                  &?&?& $0.58^{0.05}_{-0.58}$ [1]&?&? \\
Steller temperature - $T_{s}$ & not used                     & $4120^{+100}_{-160}$  [?]$^{\dagger}$ &?&?& $4400^{200}_{200}$ [1]&?&? \\
Planet radius - $R_{p}$      & $0.93^{0.07}_{-0.09}$     [1]& $0.93^{+0.07}_{-0.09}$ [1]&?&?& $0.93^{0.07}_{-0.09}$ [1]&?&? \\
Planet mass - $M_{p}$      & 1.78$\pm$0.1              [1]& 1.8$\pm$0.1 [?]$^{\dagger}$ &?&?& 1.78$\pm$0.1 [1]&?&? \\
semi-major axis - $a$       & 0.0142$\pm$0.0004         [1]& 0.0142$\pm$0.0004      [1]&?&?& 0.0142$\pm$0.0004 [1]&?&? \\
planet period - $P$       & 0.813475$\pm$1e-6         [1]& 0.813474$\pm$2e-8      [2]&?&?& .813475$\pm$1e-6 [1]&?&? \\
time of mid transit - $t_{0}$ & 2455528.86774$\pm$0.00014 [1]& not used                  &?&?& 2455528.86774$\pm$0.00014 [1]&?&? \\
orbit inclination - $i$       & $82.6^{1.3}_{-0.9}$       [1]& 82.1$\pm$0.1           [2]&?&?& $82.6^{1.3}_{-0.9}$ [1]&?&? \\
orbit eccentricity - $e$       & 0.0$\pm$0.145             [1]& not used                  &?&?& not used &?&? \\
\hline
method & pixel map & PLDAper1 - 3x3 & BLISS & BLISS  & ICA & BLISS+PFR & BLISS \\
$R_{p}/R_{s}$ & 0.15722$\pm$0.00033 & 0.159$\pm$0.001 & $0.15935^{0.00095}_{-0.0011}$ & -- & 0.1572$\pm$0.0010 & -- & 0.1589$\pm$0.0005 \\
$F_{p}/F_{s}$ [ppm] & 3739$\pm$85 & 3800$\pm$100 & 3650$\pm$140 & 3660$\pm$120 & 3870$\pm$120 & 4060$\pm$100 & 3830$\pm$80 \\
$A$ [ppm] & 1638±42.4& $1850^{+223}_{-141}$ & $1822^{+97}_{-110}$ & 1613±83 & 1800$\pm$96 & 1630±120 & 1999$\pm$62 \\
$\phi$ [deg] & 1±0.1 & 9$\pm$3 & 20.4±3.6 & 20.6±2.0 & 11.3±2.1 & 12±3 & 21±1.8 \\ \hline
[1] & \cite{hellier2011} & & & & & & \\
$[2]$ & \cite{kokori2023} & & & & & & \\
$[?]^{\dagger}$ & value not in archive & & & & & \\
? & denotes no information & & & & & \\ \hline
\end{tabular}
\\
\end{center}
\end{sidewaystable}

\begin{figure*}[h]
\centering
\hspace{-0.25in} \\
\includegraphics[scale=0.6]{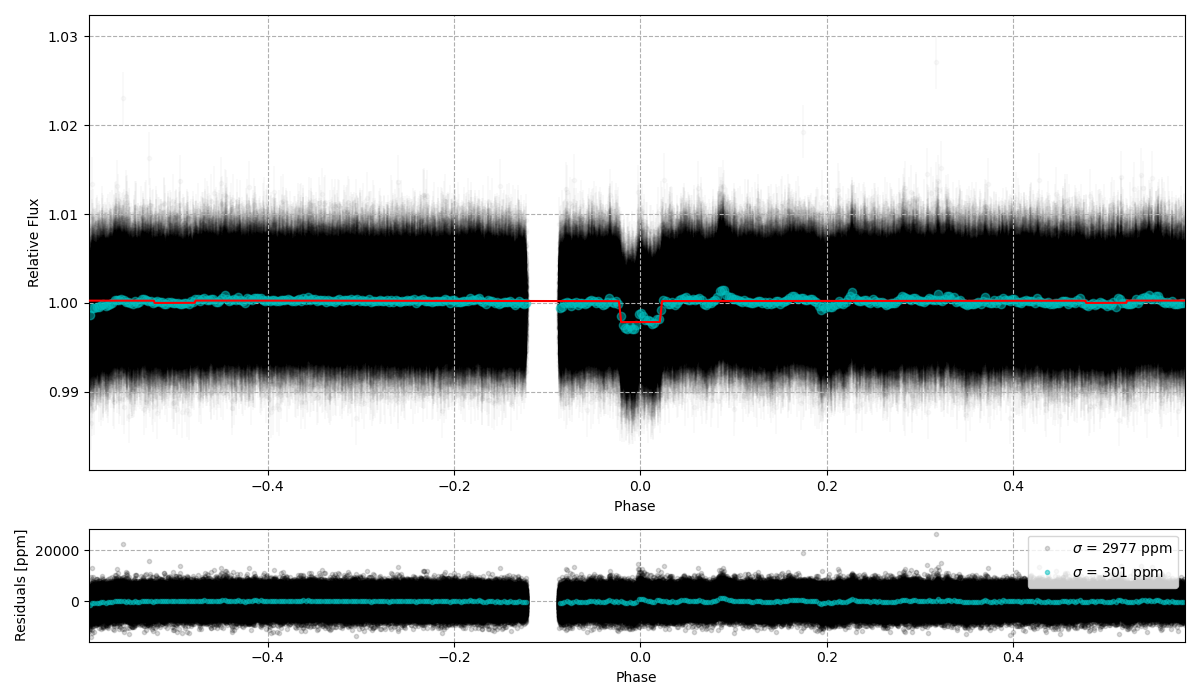}
\includegraphics[scale=0.6]{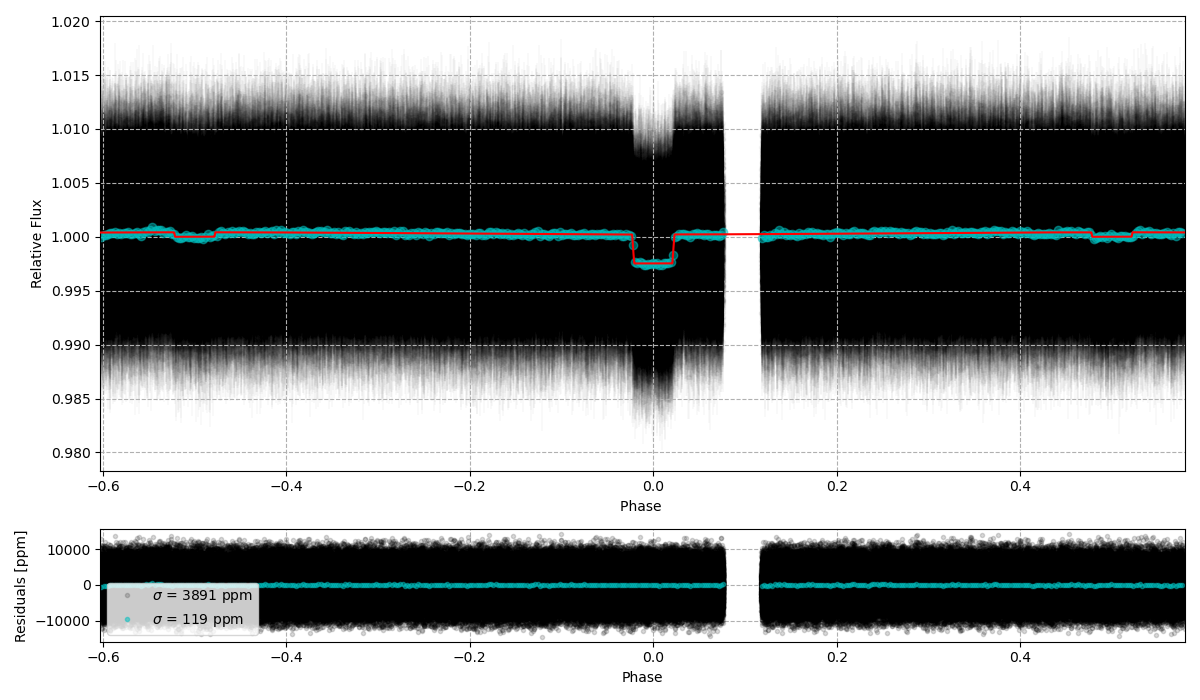}\\
\caption{HD 149026 b 3.6-$\mu$m (top) and 4.5-$\mu$m (bottom) phase curves with residual systematics clearly showing in the 3.6-$\mu$m time series. Colors same as in Figure~\ref{phasecurve}.} 
\label{fig:HD149026b}
\end{figure*}

\newpage

\section{Two Color Phase Curve Sample Parameters}

\begin{table}[h]
  \caption{Two Color Phase Curve Sample Amplitude and Phase Offset Values}
\label{tab:phasecurve_amp_phase}    
\begin{center}
\begin{tabular}{l|cc|cc} \hline \hline
 Planet & $A^{3.6 \mu m}$ & $\phi^{3.6 \mu m}$ &  $A^{4.5 \mu m}$ & $\phi^{4.5 \mu m}$ \\ 
          &  [ppm] & [degrees] & [ppm] & [degrees] \\   \hline 
HAT-P-2b &   96$\pm$0.4 &   51$\pm$0.3 &  355$\pm$26.9 &   18$\pm$1.4 \\
HAT-P-7b &  654$\pm$16.7 &   14$\pm$0.4 &  878$\pm$53.3 &   17$\pm$1.1 \\
KELT-1b &  630$\pm$19.9 &   25$\pm$0.8 &  717$\pm$51.2 &   -2$\pm$0.1 \\
KELT-14b &  428$\pm$13.2 &   51$\pm$2.2 &  869$\pm$78.5 &  -12$\pm$1.1 \\
KELT-9b &  433$\pm$1.9 &  -51$\pm$0.3 &  838$\pm$18.4 &  -27$\pm$0.6 \\
Qatar-1b & 1230$\pm$34.6 &   14$\pm$0.4 & 1132$\pm$84.3 &    9$\pm$0.7 \\
Qatar-2b &  344$\pm$14.1 &   51$\pm$2.9 & 1205$\pm$53.1 &   18$\pm$0.8 \\
WASP-12b & 1664$\pm$32.4 &   14$\pm$0.3 & 1474$\pm$73.0 &   -2$\pm$0.1 \\
WASP-121b & 1890$\pm$54.1 &   15$\pm$0.5 & 2037$\pm$76.5 &    1$\pm$0.0 \\
WASP-14b &  793$\pm$23.4 &   10$\pm$0.3 &  820$\pm$22.1 &  -11$\pm$0.3 \\
WASP-19b & 1681$\pm$60.3 &   -5$\pm$0.2 & 1608$\pm$89.1 &   -7$\pm$0.4 \\
WASP-33b & 1347$\pm$12.9 &   -4$\pm$0.1 & 1810$\pm$10.5 &  -17$\pm$0.1 \\
WASP-43b &  941$\pm$38.7 &  -25$\pm$1.1 & 1638$\pm$42.4 &    1$\pm$0.1 \\
WASP-52b & 1568$\pm$24.8 &   14$\pm$0.2 &  479$\pm$22.3 &  -51$\pm$3.3 \\
WASP-76b & 1384$\pm$15.1 &   15$\pm$0.2 & 1309$\pm$59.9 &  -11$\pm$0.5 \\
WASP-77b &  315$\pm$23.4 &  -14$\pm$1.1 &  860$\pm$24.8 &  -18$\pm$0.5 \\
\hline
\end{tabular}
\\
\end{center}
\end{table}

\begin{center}
\begin{table}
  \caption{Two Color Phase Curve Sample Planet Properties}
  \label{tab:phasecurve_planet_proterties}
\begin{tabular}{l|cccc|l} \hline \hline
Planet   & $T_{eq}(A=0)$ & Period & $R_p$ & $log(g)$  &  References\\ 
          & [K] & [days] &     [$R_{J}$] & [cm s$^{-2}$]  \\   \hline 
HAT-P-2b &
 1390$\pm$64 & 
  5.6335158$\pm$3.6e-6 & 
 1.16$^{+0.07}_{-0.06}$ & 
 4.19$^{+0.08}_{-0.09}$ &
 \cite{bonomo2017,ment2018} \\
HAT-P-7b &   
 2223$\pm$667 & 
  2.20474$\pm$1.7e-5 & 
 1.51$\pm$0.21 & 
 3.30$^{+0.15}_{-0.22}$ & 
 \cite{stassun2017} \\
KELT-1b &  
 2416$\pm$250 & 
 1.217514$\pm$1.5e-5 & 
 1.11$^{+0.03}_{-0.02}$ & 
 4.74$^{+0.03}_{-0.02}$ & 
 \cite{siverd2012} \\
KELT-14b & 
 1962$\pm$63 & 
  1.7100566$^{+3.2e-6}_{-2.6e-6}$ & 
 1.74$\pm$0.05 & 
 3.02$^{+0.03}_{-0.03}$ &
 \cite{turner2016} \\
KELT-9b  &   
4049$\pm$303 & 
  1.4811235$\pm$1.1e-6 & 
 1.89$^{+0.06}_{-0.05}$ & 
 3.30$^{+0.11}_{-0.15}$ &
 \cite{gaudi2017} \\
Qatar-1b &
 1416$\pm$58 & 
 1.4200242$\pm$2.2e-7 & 
 1.14$^{+0.03}_{-0.03}$ & 
 3.39$^{+0.03}_{-0.03}$ & 
 \cite{collins2017} \\
Qatar-2b & 
 1348$\pm$30 & 
  1.3371165$\pm$2.6e-7 & 
 1.25$\pm$0.01 & 
 3.59$^{+0.01}_{-0.01}$ & 
 \cite{mancini2014} \\
WASP-12b &   
 2504$\pm$191 & 
  1.0914225$\pm$1.4e-7 & 
 1.94$\pm$0.06 & 
 2.99$^{+0.03}_{-0.04}$ & 
\cite{stassun2017,chakrabarty2019} \\
WASP-121b&   
 2336$\pm$111 & 
  1.2749250$^{+1.5e-7}_{-1.4e-7}$ & 
 1.75$\pm$0.04 & 
 2.97$^{+0.03}_{-0.03}$ & 
 \cite{bourrier2020} \\
WASP-14b & 
 1861$\pm$308 & 
  2.2437500$\pm$1e-5 & 
 1.38$\pm$0.08 & 
 4.06$^{+0.08}_{-0.09}$ &
 \cite{stassun2017} \\
WASP-19b &  
2117$\pm$142 & 
  0.7888385$^{+7.5e-7}_{-8.2e-7}$ & 
 1.42$^{+0.04}_{-0.05}$ & 
 3.15$^{+0.04}_{-0.04}$ & 
 \cite{cortes-zuleta2020} \\
WASP-33b &   
 2781$\pm$154 & 
  1.219870$\pm$1e-6 & 
 1.59$\pm$0.07 & 
 3.31$^{+0.05}_{-0.05}$ & 
 \cite{chakrabarty2019} \\
WASP-43b &    
1379$\pm$106 & 
  0.813475$\pm$1e-6 & 
 0.93$^{+0.07}_{-0.09}$ & 
 3.71$^{+0.06}_{-0.10}$ & 
 \cite{hellier2011} \\
WASP-52b &  
 1299$\pm$42 & 
  1.7497798$\pm$1.2e-6 & 
 1.27$\pm$0.03 & 
 2.85$^{+0.03}_{-0.03}$ & 
 \cite{hebrard2013} \\
WASP-76b &    
 2182$\pm$79 & 
  1.809886$\pm$1e-6 & 
 1.83$^{+0.06}_{-0.04}$ & 
 2.83$^{+0.03}_{-0.02}$ & 
 \cite{west2016} \\
WASP-77b &  
 1691$\pm$71 & 
  1.3600285$\pm$6.2e-7 & 
 1.23$^{+0.03}_{-0.03}$ & 
 3.44$^{+0.03}_{-0.03}$ & 
 \cite{cortes-zuleta2020} \\
\hline
\end{tabular}
\\ 
\end{table}
\end{center}

\clearpage
\bibliography{ref}
\end{document}